\documentclass[journal]{IEEEtran}
\ifCLASSINFOpdf
\else
\fi

\usepackage{amsmath}
\usepackage{amssymb}
\usepackage{amsthm}
\usepackage{bm}
\usepackage[compress]{cite}

\usepackage{graphicx}
\usepackage{indentfirst}

\usepackage{subfigure}

\usepackage{lineno}
\usepackage{multirow}
\usepackage{amsfonts}
\usepackage{textcomp}
\usepackage{stfloats}

\usepackage{threeparttable}
\usepackage[bookmarksnumbered, colorlinks, citecolor=black, linkcolor=black]{hyperref}

\usepackage{float}
\usepackage{indentfirst}
\usepackage{mathrsfs}

\newtheorem{thom}{\textbf{Theorem}}
\newtheorem{asp}{\textbf{Assumption}}
\newtheorem{rek}{\textbf{Remark}}

\newtheorem{lema}{\textbf{Lemma}}

\newtheorem{pbm}{\textbf{Problem}}

\begin{document}
%
\title{Guarding a Subspace in High-Dimensional Space with Two Defenders and One Attacker}
%
%
%

\author{Rui~Yan,~\IEEEmembership{Student Member,~IEEE,}
        Zongying~Shi,~\IEEEmembership{Member,~IEEE,}
        and~Yisheng~Zhong
\thanks{This work was supported by the National Natural Science Foundation of China under Grants 61374034 and 1210012.}
\thanks{R. Yan, Z. Shi, and Y. Zhong are with the Department of Automation, Tsinghua University, Beijing 100084, China email: yr15@mails.tsinghua.edu.cn and \{szy, zys-dau\}@mail.tsinghua.edu.cn}}

\maketitle

\begin{abstract}
This paper considers a subspace guarding game in high-dimensional space which consists of a play subspace and a target subspace. Two faster defenders cooperate to protect the target subspace by capturing an attacker which strives to enter the target subspace from the play subspace without being captured. A closed-form solution is provided from the perspectives of kind and degree. Contributions of the work include the use of the attack subspace (AS) method to construct the barrier, by which the game winner can be perfectly predicted before the game starts. In addition to this inclusion, with the \emph{priori} information about the game result, a critical payoff function is designed when the defenders can win the game. Then, the optimal strategy for each player is explicitly reformulated as a saddle-point equilibrium. Finally, we apply these theoretical results to a half-space guarding game in three-dimensional space. Since the whole achieved developments are analytical, they require a little memory without computational burden and allow for real-time updates, beyond the capacity of traditional Hamilton-Jacobi-Isaacs method. It is worth noting that this is the first time in the current work to consider the target guarding games for arbitrary high-dimensional space, and in a fully analytical form.
\end{abstract}


%
\IEEEpeerreviewmaketitle

\section{Introduction}\label{introsection}
\IEEEPARstart{M}{ultiplayer} differential games have been studied extensively, and present important and interesting, but also challenging, problems in robotics, aircraft control, security, and other domains \cite{Ho1098197,petrosjan1993differential,engwerda2005lq,Ba1999Dynamic,Liu6907859,Dong7862774}. In these types of problems multiplayer reach-avoid differential games attract numerous attention recently \cite{Mitchell2005A,Margellos2011Ham,Fisac2015Reach,Barron2017Reach}, as they can provide guarantees of safety and meantime goal satisfactions from the perspective of game theory. Reach-avoid differential games consider the scenario of one or more defenders trying to maneuver and protect a predefined target by reaching a relatively small distance to one or more attackers, which strive to hit the target and meanwhile keep a safe distance from the defenders before the arrival. Differential games of this setup are also called target guarding games \cite{Mohanan2018Toward}.

Such differential games encompass a large number of realistic adversarial situations. For instance, in \cite{ChenMo2016Multiplayer} multiple defenders in a planar domain are used to prevent multiple attackers from reaching a static target set. The work by Garcia \emph{et al.} \cite{Garcia8340791} studied the dynamic game of an attacker pursuing a target aircraft protected by a defender, and the associated state space dimension is six. A multi-agent collision avoidance problem was considered in \cite{Mylvaganam7909033}, in which each agent is steered from its initial position to a desired goal while avoiding collisions with obstacles and other agents. In \cite{Mohanan2018Toward}, the target guarding problem \cite{Is1967Diff} was revisited to investigate the real-time implementation for the optimal solution. The linear quadratic game theory was employed to solve the problem of defending an asset in \cite{Li5751240}. Motivated by football game, the authors \cite{YanReaTwo2018} proposed a three-player differential game in a square region where two defenders attempt to capture an attacker before it reaches a specified edge of the game boundary, which involves six states. 

As reach-avoid games, or called target guarding games, are prevalent in many engineering applications, many methods have been proposed to deal with them, and have enjoyed great success in certain conditions. In general, the Hamilton-Jacobi (HJ) reachability is a powerful tool, as it can compute the backward reachable set, defined as the set of states from which a system is guaranteed to have a control strategy to reach a target set of states, regardless of disturbances and antagonistic controls. However, the computation of this method relies on gridding, and hence suffers from the well-known curse of dimensionality. Actually, the standard HJ method can only efficiently handle with the systems of up to five states \cite{Chen8267187}. Attempts to circumvent this problem, via symmetry of the considered systems \cite{Maidens2018Exploiting}, linear programming \cite{Kariotoglou2017The}, system decomposition \cite{Mo7989015}, and boundary analysis \cite{Xue2017Rea}, have been applied to systems with special structures, but not scalable to larger problems and still subjected to the inherent inaccuracy.

For special problem setups and system dynamics, geometric control reveals a huge capacity of providing strategies for the players. For example, Voronoi diagrams are employed to deal with group pursuit of one or more evaders, such as minimizing the area of generalized Voronoi partition of the evader \cite{ZHOU201664,Pierson2017Int}, or pursuing the evader in a relay way \cite{Bakolas2012Relay}. Specifically, Apollonius circle is introduced to analyze the capture of high-speed evaders \cite{Ramana2016Pursuit}, with a better performance than Voronoi-based approaches. For more complex game domains, such as in the presence of obstacles, Euclidean shortest path is employed to construct the dominance region for each player \cite{Oyler2016Pursuit}. More recently, paths of defense have been designed to approximate the reach-avoid set in \cite{ChenMo2016Multiplayer}.

With advances in computation speed, model-predictive control and reinforcement learning methods have also been used. As discussed in \cite{Eklund2012Switched}, a supervisory controller based on model-predictive control was designed and tested in the switched and symmetric pursuit evasion games. In \cite{Polak2016Method}, a feedback, receding horizon control law was proposed for the defenders to guard a harbor. Raslan \emph{et al.} \cite{Raslan7490540} combined a fuzzy logic controller with reinforcement learning to train an invader for the guarding a territory game.

The subspace guarding game addressed in this paper involves two defenders and one attacker moving in the game space $\mathbb{R}^n(n\ge2)$, with each player having $n$ states. The game space $\mathbb{R}^n$ is divided into two subspaces by a hyperplane. The attacker initially lying in one of two subspaces, attempts to enter the other subspace by penetrating the splitting hyperplane, while two faster defenders aim at protecting the latter subspace and strive to prevent the attacker by capturing it. From another side, this game can also be viewed as an evader (the attacker) tries to escape from a subspace through its boundary which is a hyperplane, while avoiding adversaries and moving obstacles formulated as a defense team, especially, two opponents are considered. To the authors' knowledge, this is a first attempt to address target guarding games in high-dimensional space. The current works involving differential games focus on no more than three dimensional game space \cite{Shinar1102372,Li7482692,HAYOUN2017122,Bopardikar20112067}.

Traditionally, the core for reach-avoid games is to compute the boundary of the reach-avoid set, also called barrier \cite{Is1967Diff}, by which the whole state space is split into two disjoint subspaces: defender winning subspace (DWS) and attacker winning subspace (AWS). The DWS is the set of initial states, from which the defenders are capable of guaranteeing the attacker's capture before it reaches the target subspace. The set of initial states which lead to the attacker's successful entry into the target subspace without being captured, is the AWS. Since the barrier plays a crucial role in determining the game winner ahead of time, several methods have been proposed to construct it, such as HJI method, geometrical arguments and numerical approximation \cite{merz1971homicidal,getz1981two,ruiz2016differential,Sun2015Pur,Lsler5354443,Zha2016Con,Bhattacharya2016,MACIAS2018271}.

The main contributions are as follows. First, for the game of kind \cite{Is1967Diff}, an attack subspace (AS) method is proposed to construct the barrier analytically, and this is the first time in the existing literature to construct the barrier directly for arbitrarily high-dimensional space. Moreover, since the constructed barrier is represented in a closed form, the high computational complexity or inaccuracy arising in other methods introduced above, are overcome. Thus, this method is applicable for real-time implementation. Second, the solution to a game of degree \cite{Is1967Diff} is provided. In view of the guaranteed winning for the defenders, a practical payoff function is designed, and the optimal strategy for each player, which essentially is a saddle-point equilibrium, is elaborated and referred as a command to an optimal point which is exactly located. The work \cite{YanReaTwo2018} is most similar to this work in its approach. However, our current work not only constructs the barrier analytically for guarding a subspace of arbitrary dimension, but also presents all optimal strategies in an analytical form. 

The rest of this paper is organized as follows. In Section \ref{problemdescriptionsection}, we formulate the game, and state our assumptions, notations and problems. Section \ref{simplifications} performs efficient simplification. In Section \ref{sectionthreeA}, several important concepts and properties are presented. In Sections \ref{subsectionb1v1} and \ref{subsectionb2v1}, the expressions of the barrier and winning subspaces are derived. In Section \ref{Optimal Strategies}, a game of degree is investigated, and the related saddle-point equilibrium is discussed in detail. Section \ref{threeexample} provides a three-dimensional example to highlight the theoretical developments. Finally, Section \ref{conclusion} concludes the paper.

\begin{figure}\centering
\graphicspath{{figures/}}
\includegraphics[width=70mm,height=47mm]{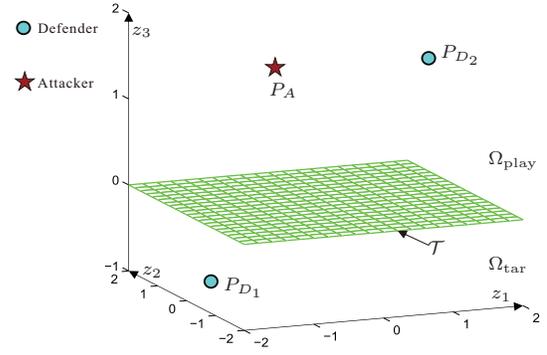}
\put(-20,70){$\scriptstyle{\Omega_{\rm play}}$}
\put(-20,30){$\scriptstyle{\Omega_{\rm tar}}$}
\put(-120,22){$\scriptstyle{P_{D_1}}$}
\put(-38,110){$\scriptstyle{P_{D_2}}$}
\put(-103,96){$\scriptstyle{P_{A}}$}
\put(-19,19){$\scriptstyle{z_1}$}
\put(-151,28){$\scriptstyle{z_2}$}
\put(-155,119){$\scriptstyle{z_3}$}
\put(-43,36){$\scriptstyle{\mathcal{T}}$}
\caption{Subspace guarding game with two defenders and one attacker in three-dimensional space ($n=3$), where $\mathcal{T}=\{\bm{z}\in\mathbb{R}^3|z_3=0\},\Omega_{\rm tar}=\{\bm{z}\in\mathbb{R}^3|z_3\leq0\}$ and $\Omega_{\rm play}=\{\bm{z}\in\mathbb{R}^3|z_3>0\}$.\label{fig:game}}
\end{figure}

\section{Problem Description}\label{problemdescriptionsection}
A differential game of guarding a subspace in high-dimensional space is considered. The game is played in $\mathbb{R}^n$ $(n\in N,n\ge2,\mathbb{R}^n=\mathbb{R}^{n\times1})$, in which a hyperplane $\mathcal{T}$ splits the game space $\mathbb{R}^n$ into two disjoint subspaces $\Omega_{\rm tar}$ and $\Omega_{\rm play}$, and their expressions are given as follows:
\begin{equation}\begin{aligned}\label{ori_tar}
&\mathcal{T}=\{\bm{z}\in\mathbb{R}^n|K^\mathsf{T}\bm{z}=b\},\Omega_{\rm{tar}}=\{\bm{z}\in\mathbb{R}^n|K^\mathsf{T}\bm{z}\leq b\}\\
&\Omega_{\rm{play}}=\{\bm{z}\in\mathbb{R}^n|K^\mathsf{T}\bm{z}>b\}
\end{aligned}\end{equation}
where $K\in\mathbb{R}^n$ and $b\in\mathbb{R}$ are the known parameters, and $K$ is a nonzero vector. Note that $\mathcal{T}\subset\Omega_{\rm tar}$. Two defenders $P_{D_1},P_{D_2}$ and one attacker $P_A$, assumed to be three mass points in $\mathbb{R}^n$, can move freely with simple motion \cite{Is1967Diff}, i.e., they are able to change the directions of their motion at each instant of time. The attacker $P_A$ is considered to have been captured as soon as his Euclidean distance from the closer defender becomes equal to zero. The attacker, starting from $\Omega_{\rm{play}}$, aims at reaching $\Omega_{\rm{tar}}$ without being captured, while two defenders, initially distributed in any positions of the game space, cooperate to guard $\Omega_{\rm{tar}}$ by capturing $P_A$. Thus, these two subspaces $\Omega_{\rm{tar}}$ and $\Omega_{\rm{play}}$ shall be called \emph{target subspace} and \emph{play subspace} respectively. We call $\mathcal{T}$ as \emph{target hyperplane (TH)}. If $n=3$, the game components are shown in Fig. \ref{fig:game}.

The game terminates when the attacker reaches the target subspace $\Omega_{\rm{tar}}$ before captured, or at least one of two defenders captures the attacker in $\Omega_{\rm{play}}$. If the former happens, the attacker wins, and if the latter happens, two defenders win.

Define the unit control set $\mathcal{U}=\{\bm{u}\in\mathbb{R}^n|\|\bm{u}\|_2=1\}$, where $\|\cdot\|_2$ stands for the Euclidean norm in $\mathbb{R}^n$. Denote the positions, or called states, of $P_{D_i}$ and $P_A$ at time $t$ in $\mathbb{R}^n$ by $\bm{x}_{D_i}(t)=(x_{D_i,1}(t),\cdots,x_{D_i,n}(t))^\mathsf{T}$ and $\bm{x}_A(t)=(x_{A,1}(t),\cdots,x_{A,n}(t))^\mathsf{T}$ respectively, where $\mathsf{T}$ stands for the transposition. The dynamics of three players for $t\ge0$ have the form
 \begin{equation}\begin{aligned}\label{dynamics}
\dot{\bm{x}}_{D_i}(t)&=v_{D}\bm{d}_i(t),&\bm{x}_{D_i}(0)&=\bm{x}_{D_i}^0,i=1,2\\
\dot{\bm{x}}_A(t)&=v_A\bm{a}(t),&\bm{x}_A(0)&=\bm{x}_A^0.
\end{aligned}\end{equation}
 Here, $\bm{x}_{D_i}^0=(x_{D_i,1}^0,\cdots,x_{D_i,n}^0)^\mathsf{T}$ is the initial position of $P_{D_i}$, $\bm{x}_A^0=(x_{A,1}^0,\cdots,x_{A,n}^0)^\mathsf{T}$ is the initial position of $P_A$, and the control inputs at time $t$ for $P_{D_i}$ and $P_A$ are their respective instantaneous unit headings $\bm{d}_i(t)\in\mathcal{U}$ and $\bm{a}(t)\in\mathcal{U}$. The positive parameters $v_{D}$ and $v_A$ are the speeds of $P_{D_i}$ and $P_A$ respectively. Thus, the whole state space is $\mathbb{R}^{3n}$. Unless for clarity, for simplicity, $t$ will be omitted hereinafter.
 
Note that two defenders are homogeneous, as they have the same speed $v_D$. Let $\mathcal{U}^2=\mathcal{U}\times\mathcal{U}$ denote the joint control set of two defenders, and since all possible cooperations between them are considered, the controls $\bm{d}_1$ and $\bm{d}_2$ will be selected simultaneously. Thus, let $\bm{d}=[\bm{d}_1^\mathsf{T},\bm{d}_2^\mathsf{T}]^\mathsf{T}\in\mathcal{U}^2$ denote the control of the defense team which can be regarded as a single player with two inputs.
\subsection{Information Structure, Strategy and Payoffs}
We focus on a non-anticipative information structure, as commonly adopted in the differential game literature (see for example,  \cite{Mitchell2005A},\cite{Elliott1972The}). Under this information structure, two defenders are allowed to make decisions about their current inputs with all the information of the speeds and current positions of all players, plus the attacker's current input. While the attacker is at a slight disadvantage under this information structure, at a minimum he has access to sufficient information to use the speeds and current positions of all players, because the defenders must declare their strategies before the attacker chooses a specific input and thus the attacker can determine the response of the defenders to any input signal. Hence, the target guarding games or called reach-avoid games formulated here are an instantiation of the Stackelberg game \cite{Ba1999Dynamic}.

Determining which team will win the game gives rise to a game of kind \cite{Is1967Diff}, which provides two outcomes of the game depending on which team can achieve its objective. In practice, with the \emph{prior} game results extracted out from the game of kind, it is a natural instinct to investigate the optimal strategies for the players in their winning subspaces by designing some critical payoff functions.

If the initial state occurs in the part of state space $\mathbb{R}^{3n}$ where two defenders can guarantee their winning, assume that two defenders want to capture the attacker at a point with the maximum distance to $\Omega_{\rm{tar}}$, while the attacker tries to minimize its final distance to $\Omega_{\rm{tar}}$ although the capture cannot be avoided. Thus, this payoff function, also called attacker-target terminal distance, is given as follows:
\begin{equation}\label{DDSpf}
J_{\mathcal{T}}(\bm{d},\bm{a};\bm{x}_{D_1}^0,\bm{x}_{D_2}^0,\bm{x}_A^0)=\min_{\bm{p}\in\mathcal{T}}\|\bm{x}_A(t_1)-\bm{p}\|_2
\end{equation}
where $t_1$ is the time when $P_A$ is captured.

If the initial state lies in the part of state space $\mathbb{R}^{3n}$ where the solution of the game of kind indicates that, under optimal plays, the attacker will reach $\Omega_{\rm{tar}}$ without being captured by two defenders, the following payoff function is proposed:
\begin{equation}\label{ADSpf}
J_{d}(\bm{d},\bm{a};\bm{x}_{D_1}^0,\bm{x}_{D_2}^0,\bm{x}_A^0)=\min_{i=1,2}\|\bm{x}_{D_i}(t_2)-\bm{x}_A(t_2)\|_2
\end{equation}
where $t_2$ is the attacker's first arrival time into $\Omega_{\rm tar}$. This payoff function (\ref{ADSpf}), also called defender-attacker terminal distance or safe distance on arrival, can be interpreted that: The attacker strives to maximize its distance from the closer defender when arriving at $\mathcal{T}$, while two defenders seek the opposite. Moreover, this payoff function also indicates that the attacker wants to reach $\mathcal{T}$ under the safest condition. We emphasize here that the payoff function \eqref{ADSpf} is introduced just for the later proof. 

\subsection{Assumptions and Notations}\label{notations}
We summarize here the assumptions we shall need throughout the paper. The first ones are concerned with the initial configurations of three players.
\begin{asp}\label{aspisolation}
$\|\bm{x}_{D_1}^0-\bm{x}_{D_2}^0\|_2>0,\|\bm{x}_{D_i}^0-\bm{x}_A^0\|_2> l,i=1,2$.
\end{asp}
\begin{asp}\label{asprelaxed}\rm
$K^\mathsf{T}\bm{x}_A^0>b,\bm{x}_{D_i}^0\in\mathbb{R}^n$, that is, $\bm{x}_A^0\in\Omega_{\rm play},\bm{x}_{D_i}^0\in\Omega_{\rm play}\cup\Omega_{\rm tar}, i=1,2$.
\end{asp}

The explanations for these two assumptions are as follows. \autoref{aspisolation} states that all players start the game from different initial positions and $P_A$ is not captured by two defenders initially. Since the point-capture case is considered, we set $l=0$. Note that \autoref{asprelaxed} confines $\bm{x}_A^0$ in $\Omega_{\rm{play}}$, which comes from our game setup.

Define $\alpha=v_A/v_{D}$ to be the speed ratio between $P_A$ and $P_{D_i}$. We focus on faster defenders in this paper.
\begin{asp}\label{aspratio}\rm
$v_{D}>v_A>0$, i.e., $0<\alpha<1$.
\end{asp}

Let $\bm{e}_i\in\mathbb{R}^{n}$ denote the vector of its $i$th element equal to 1 and the others equal to 0, $I_n$ denote the identity matrix of size $n$, and $0_{m\times n}$ denote the $m\times n$ zero matrix. For any $\bm{z}\in\mathbb{R}^n$, denote the remaining part when its $i$th element is removed by $\bm{z}_{-i}\in\mathbb{R}^{n-1}$. For example, $\bm{x}_{D_i,-n}^0\in\mathbb{R}^{n-1}$ denotes the remaining part of $\bm{x}_{D_i}^0$ when its $n$-th element $x_{D_i,n}^0$ is removed.

For clarity, introduce several critical notations related to the initial positions of two defenders. Define $A_{ij}=\bm{x}_{D_i,-n}^0-\bm{x}_{D_j,-n}^0,B_{ij}=(\bm{x}_{D_i,-n}^0+\bm{x}_{D_j,-n}^0)/2,C_{ij}=\|A_{ij}\|_2^2I_{n-1}-A_{ij}A_{ij}^\mathsf{T},m_{ij}=x_{D_i,n}^0-x_{D_j,n}^0$, and $w_{ij}=(\|\bm{x}_{D_i}^0\|_2^2-\|\bm{x}_{D_j}^0\|_2^2)/2$. Thus, it can be seen that $A_{ij}\in\mathbb{R}^{n-1}$, $B_{ij}\in\mathbb{R}^{n-1}$, $C_{ij}\in\mathbb{R}^{(n-1)\times(n-1)}$, $m_{ij}\in\mathbb{R}$ and 
$w_{ij}\in\mathbb{R}$. Also note that $B_{ij}=B_{ji}$ and $C_{ij}=C_{ji}$.

Define the following two matrixes, which will be used for characterizing the barrier and winning subspaces 
\begin{equation}\begin{aligned}\label{Ximatrix1122}
\Xi_i&=\begin{bmatrix}
\begin{smallmatrix}
-I_{n-1} & 0_{(n-1)\times1} &  \bm{x}_{D_i,-n}^0\\
0_{1\times(n-1)} & 1/\alpha^2-1 & 0\\
(\bm{x}_{D_i,-n}^0)^\mathsf{T} & 0 & \alpha^2(x_{D_i,n}^0)^2-\|\bm{x}_{D_i}^0\|_2^2\\
\end{smallmatrix}
\end{bmatrix}\\
\Xi_{ij}&=\begin{bmatrix}
\begin{smallmatrix}
-\zeta_{ij,2} & 0_{(n-1)\times1} & -\zeta_{ij,3}\\
0_{1\times(n-1)} & \zeta_{ij,1}& 0\\
-\zeta_{ij,3}^\mathsf{T} & 0 & -\zeta_{ij,4}\\
\end{smallmatrix}
\end{bmatrix}
\end{aligned}\end{equation}
where the involved parameters are given as follows:
\begin{equation}\begin{aligned}\label{zetadefinition}
&\zeta_{ij,1}=(1-\alpha^2)\|A_{ij}\|_2^2,\zeta_{ij,2}=C_{ij}-(1-\alpha^2)\|A_{ij}\|_2^2I_{n-1}\\
&\zeta_{ij,3}=(1-\alpha^2)A_{ij}w_{ij}-\alpha^2C_{ij}B_{ij}\\
&\zeta_{ij,4}=(1-\alpha^2)\alpha^2A_{ij}^\mathsf{T}(\|\bm{x}_{D_j}^0\|_2^2\bm{x}_{D_i,-n}^0-\|\bm{x}_{D_i}^0\|_2^2\bm{x}_{D_j,-n}^0)\\
&\qquad\qquad\qquad\qquad\qquad +\alpha^4B_{ij}^\mathsf{T}C_{ij}B_{ij}-(1-\alpha^2)^2w_{ij}^2.
\end{aligned}\end{equation}
Notice that $\zeta_{ij,1}$ and $\zeta_{ij,4}$ are scalar, $\zeta_{ij,2}\in\mathbb{R}^{(n-1)\times(n-1)}$ and $\zeta_{ij,3}\in\mathbb{R}^{n-1}$. Additionally, $\Xi_{i}$ and $\Xi_{ij}$ are two $(n+1)\times(n+1)$ matrixes. It can be verified that $\Xi_{ij}=\Xi_{ji}$.

Let $\mathcal{N}_1,\mathcal{N}_{12}$ and $\mathcal{N}_2$ denote three subspaces of TH $\mathcal{T}$, each position of which $P_{D_1}$ can reach with less, the same and more minimal time with respect to $P_{D_2}$, respectively. Thus, they can be mathematically formulated as follows:
\begin{equation}\begin{aligned}\label{subspacesofTS1}
\mathcal{N}_1&=\{\bm{z}\in\mathbb{R}^n|\|\bm{z}-\bm{x}_{D_1}^0\|_2<\|\bm{z}-\bm{x}_{D_2}^0\|_2,K^\mathsf{T}\bm{z}=b\}\\
\mathcal{N}_{12}&=\{\bm{z}\in\mathbb{R}^n|\|\bm{z}-\bm{x}_{D_1}^0\|_2=\|\bm{z}-\bm{x}_{D_2}^0\|_2,K^\mathsf{T}\bm{z}=b\}\\
\mathcal{N}_2&=\{\bm{z}\in\mathbb{R}^n|\|\bm{z}-\bm{x}_{D_1}^0\|_2>\|\bm{z}-\bm{x}_{D_2}^0\|_2,K^\mathsf{T}\bm{z}=b\}
\end{aligned}\end{equation}
which will play a crucial role in our following analysis.

Next, define two notations associated with the relative initial position of $P_A$ with respect to the defender $P_{D_i}$ as follows:\begin{equation}\label{thetaanddelta}
\theta_i=\frac{\bm{x}_A^0-\alpha^2\bm{x}_{D_i}^0}{1-\alpha^2},\delta_i=\frac{\alpha\|\bm{x}_A^0-\bm{x}_{D_i}^0\|_2}{1-\alpha^2}
\end{equation}
whose geometric meanings will be stated in Section \ref{sectionthreeA}. Note that $\theta_i\in\mathbb{R}^n$ and $\delta_i\in\mathbb{R}$.

Finally, introduce several notations to simplify the expressions of optimal strategies obtained later. Define 
\begin{equation}\begin{aligned}\label{R1R2R3}
&R_1=A_{12}A_{12}^\mathsf{T}+m_{12}^2I_{n-1},R_2=(\theta_{1,n}m_{12}-w_{12})A_{12}\\
&-m_{12}^2\theta_{1,-n},r_3=w_{12}^2-2\theta_{1,n}w_{12}m_{12}+m_{12}^2(\|\theta_1\|_2^2-\delta_1^2).
\end{aligned}\end{equation}
Notice that $R_1\in\mathbb{R}^{(n-1)\times(n-1)}$, $R_2\in\mathbb{R}^{n-1}$ and $r_3\in\mathbb{R}$. The reasons why (\ref{R1R2R3}) is used will be explained in Section \ref{Optimal Strategies}.

\subsection{Problems}
For this subspace guarding game in high-dimensional space, two problems will be addressed. 
\begin{pbm}[Game of kind]\label{kind}
Given $K,b$, and any admissible initial configuration $K^\mathsf{T}\bm{x}_A^0>b$ and $\bm{x}_{D_i}^0\in\mathbb{R}^n(i=1,2)$, which team can guarantee its own winning? Does this game end up with a successful capture or a successful attack when both team adopt their optimal strategies?
\end{pbm}
\begin{pbm}[Game of degree]\label{degreeDDS}
If the defense team can guarantee the capture in $\Omega_{\rm play}$, two defenders or the attacker needs to find a proper control input based on their or its accessible information, such that the maximum or minimum of the payoff function (\ref{DDSpf}) is achieved. In other words, find a saddle-point equilibrium $(\bm{d}^*,\bm{a}^*)$ of the maxmin problem:
\begin{equation}\label{DDSvf}
V_\mathcal{T}^2(\bm{x}_{D_1}^0,\bm{x}_{D_2}^0,\bm{x}_A^0)=\max_{\bm{d}\in\mathcal{U}^2}\min_{\bm{a}\in\mathcal{U}}J_{\mathcal{T}}(\bm{d},\bm{a};\bm{x}_{D_1}^0,\bm{x}_{D_2}^0,\bm{x}_A^0)
\end{equation} 
where $V_\mathcal{T}^2$ is the corresponding value function.
\end{pbm}

\section{Efficient Simplification}\label{simplifications}
In this section, we describe this game in a clearer way. The TH and two subspaces in (\ref{ori_tar}) can be represented by
\begin{equation}\begin{aligned}\label{new_tar}
&\mathcal{T}=\{\bm{z}\in\mathbb{R}^n|z_n=0\},\Omega_{\rm{tar}}=\{\bm{z}\in\mathbb{R}^n|z_n\leq0\}\\
&\Omega_{\rm{play}}=\{\bm{z}\in\mathbb{R}^n|z_n>0\}.
\end{aligned}\end{equation}
Thus, without loss of generality, we focus on (\ref{new_tar}) hereinafter.

Hence, (\ref{new_tar}) implies that the three subspaces $\mathcal{N}_1,\mathcal{N}_{12}$ and $\mathcal{N}_2$ defined in (\ref{subspacesofTS1}) can be rewritten in the new coordinate system as follows. For $\mathcal{N}_1$, since $K^\mathsf{T}\bm{z}=b\Leftrightarrow z_n=0$, then
\begin{equation*}\begin{aligned}\label{subspacesofTS23}
&\|\bm{z}-\bm{x}_{D_1}^0\|_2<\|\bm{z}-\bm{x}_{D_2}^0\|_2\Rightarrow2(\bm{x}_{D_1,-n}^0-\bm{x}_{D_2,-n}^0)^{\mathsf{T}}\bm{z}_{-n}\\
&>\|\bm{x}_{D_1}^0\|_2^2-\|\bm{x}_{D_2}^0\|_2^2\Rightarrow A_{12}^{\mathsf{T}}\bm{z}_{-n}>w_{12}.
\end{aligned}\end{equation*}
Therefore, in the similar way, these three subspaces in the new coordinate system are given as follows:
\begin{equation}\begin{aligned}\label{subspacesofTS}
\mathcal{N}_1&=\{\bm{z}\in\mathbb{R}^n|A_{12}^\mathsf{T}\bm{z}_{-n}>w_{12},z_n=0\}\\
\mathcal{N}_{12}&=\{\bm{z}\in\mathbb{R}^n|A_{12}^\mathsf{T}\bm{z}_{-n}=w_{12},z_n=0\}\\
\mathcal{N}_2&=\{\bm{z}\in\mathbb{R}^n|A_{21}^\mathsf{T}\bm{z}_{-n}>w_{21},z_n=0\}.
\end{aligned}\end{equation}

To simplify the analysis further, \autoref{kind} can be reformulated: By fixing two defenders' initial positions, we aim to find the subspace of $\Omega_{\rm play}$ where if the attacker initially lies, two defenders can guarantee the capture before the attacker reaches the TH $\mathcal{T}$, which is the DWS, and find the subspace of $\Omega_{\rm play}$ allowing for a successful attacking strategy for the attacker, which is the AWS. The surface that separates these two subspaces is the barrier. Fixing two defenders' initial positions provides a clear illustration of the barrier and thus two wining subspaces as the functions of these initial positions.

Let $\mathcal{B}^1(\bm{x}_{D_i}^0,\alpha),\mathcal{W}^1_D(\bm{x}_{D_i}^0,\alpha)$ and $\mathcal{W}^1_A(\bm{x}_{D_i}^0,\alpha)$ denote the barrier, DWS and AWS determined by $P_{D_i}$ respectively, which implies that these variants only depend on $P_{D_i}$'s initial position and the speed ratio as is proved below.

For two defenders $P_{D_1},P_{D_2}$ and one attacker $P_A$, let $\mathcal{B}^2(\bm{x}_{D_1}^0,\bm{x}_{D_2}^0,\alpha),\mathcal{W}^2_D(\bm{x}_{D_1}^0,\bm{x}_{D_2}^0,\alpha)$ and $\mathcal{W}^2_A(\bm{x}_{D_1}^0,\bm{x}_{D_2}^0,\alpha)$ denote  the associated barrier, DWS and AWS respectively.

If $P_{D_i}$ satisfies $\bm{x}_{D_i}^0\in\Omega_{\rm{tar}}$, we introduce a virtual defender $\tilde{P}_{D_i}$ with initial position $\tilde{\bm{x}}_{D_i}^0$ such that $\tilde{x}_{D_i,m}^0=x_{D_i,m}^0(m=1,\cdots,n-1)$ and $\tilde{x}_{D_i,n}^0=-x_{D_i,n}^0$. Thus, it can be easily observed that the virtual defender and its original defender are symmetric with respect to the TH $\mathcal{T}$. 

Next, a property on the barrier construction is stated.
\begin{lema}[Mirror property]\label{mirror}\rm
For $P_{D_i}$ and $j\neq i(i,j\in\{1,2\})$, if $\bm{x}_{D_i}^0\in\Omega_{\rm{tar}}$, $\mathcal{B}^2(\bm{x}_{D_i}^0,\bm{x}_{D_j}^0,\alpha)=\mathcal{B}^2(\tilde{\bm{x}}_{D_i}^0,\bm{x}_{D_j}^0,\alpha)$ holds.

\emph{Proof:} We postpone the proof to Appendix.
\qed
\end{lema}
\begin{rek}\label{mirrorrek}\rm
Note that \autoref{mirror} shows that the virtual defender plays the same role with its original defender in barrier construction and thus in determining winning subspaces. Hence, for clarity, all proofs below involving the barrier construction will only focus on the discussion under the condition $\bm{x}_{D_i}^0\in\Omega_{\rm play}\cup\mathcal{T}(i=1,2)$, but these relevant results are stated and hold under \autoref{asprelaxed}. 
\end{rek}

\section{Barrier and Winning Subspaces}
This section focuses on \autoref{kind}, namely, which team will win the game, which is a game of kind.

\subsection{Preliminaries}\label{sectionthreeA}
We begin our analysis with some preliminary results. Let the set of points in the game space which $P_A$ can reach before the defender(s), regardless of the defender(s)' best effort, be called AS, and the surface which bounds AS is called \emph{the boundary of AS (BAS)}. 

Denote the AS and BAS associated with $P_{D_i}$ and $P_A$ by $\mathcal{R}_A^1(\bm{x}_A^0,\bm{x}_{D_i}^0,\alpha)$ and ${\rm bas}^1(\bm{x}_A^0,\bm{x}_{D_i}^0,\alpha)$ respectively. Thus, according to the definitions, the AS and BAS can be given by\begin{equation}\label{AR1}\begin{aligned}
\mathcal{R}_A^1(\bm{x}_A^0,\bm{x}_{D_i}^0,\alpha)&=\{\bm{z}\in\mathbb{R}^n|\|\bm{z}-\bm{x}_A^0\|_2<\alpha\|\bm{z}-\bm{x}_{D_i}^0\|_2\}\\
{\rm{bas}}^1(\bm{x}_A^0,\bm{x}_{D_i}^0,\alpha)&=\{\bm{z}\in\mathbb{R}^n|\|\bm{z}-\bm{x}_A^0\|_2=\alpha\|\bm{z}-\bm{x}_{D_i}^0\|_2\}.
\end{aligned}\end{equation}
Note that 
\begin{equation*}\label{ARtransfer1}\begin{aligned}
&\|\bm{z}-\bm{x}_A^0\|_2<\alpha\|\bm{z}-\bm{x}_{D_i}^0\|_2\Rightarrow\\
&(1-\alpha^2)\|\bm{z}\|_2^2-2(\bm{x}_A^0-\alpha^2\bm{x}_{D_i}^0)^\mathsf{T}\bm{z}<\alpha^2\|\bm{x}_{D_i}^0\|_2^2-\|\bm{x}_A^0\|_2^2\\
&\Rightarrow\Big\|\bm{z}-\frac{\bm{x}_A^0-\alpha^2\bm{x}_{D_i}^0}{1-\alpha^2}\Big\|_2^2<\frac{\alpha^2\|\bm{x}_A^0-\bm{x}_{D_i}^0\|_2^2}{(1-\alpha^2)^2}\\
&\Rightarrow\|\bm{z}-\theta_i\|_2^2<\delta_i^2
\end{aligned}\end{equation*}
where $\theta_i$ and $\delta_i$ are defined in (\ref{thetaanddelta}). Thus, $\mathcal{R}_A^1(\bm{x}_A^0,\bm{x}_{D_i}^0,\alpha)$ is the interior of a ball of radius $\delta_i$ centered at $\theta_i$, which also explains the geometric meanings of $\delta_i$ and $\theta_i$. Naturally, ${\rm bas}^1(\bm{x}_A^0,\bm{x}_{D_i}^0,\alpha)$ is the sphere. Then, the AS $\mathcal{R}_A^1(\bm{x}_A^0,\bm{x}_{D_i}^0,\alpha)$ and BAS ${\rm{bas}}^1(\bm{x}_A^0,\bm{x}_{D_i}^0,\alpha)$ in (\ref{AR1}) can be simplified:
\begin{equation}\label{ARtransfer112}\begin{aligned}
\mathcal{R}_A^1(\bm{x}_A^0,\bm{x}_{D_i}^0,\alpha)&=\{\bm{z}\in\mathbb{R}^n|\|\bm{z}-\theta_i\|_2<\delta_i\}\\
{\rm{bas}}^1(\bm{x}_A^0,\bm{x}_{D_i}^0,\alpha)&=\{\bm{z}\in\mathbb{R}^n|\|\bm{z}-\theta_i\|_2=\delta_i\}.
\end{aligned}\end{equation}
If $n=2$, then ${\rm{bas}}^1(\bm{x}_A^0,\bm{x}_{D_i}^0,\alpha)$ is actually the Apollonius circle \cite{Ramana2016Pursuit}, and $\mathcal{R}_A^1(\bm{x}_A^0,\bm{x}_{D_i}^0,\alpha)$ is the interior of this circle.

Let $\mathcal{R}_A^2(\bm{x}_A^0,\bm{x}_{D_1}^0,\bm{x}_{D_2}^0,\alpha)$ and ${\rm bas}^2(\bm{x}_A^0,\bm{x}_{D_1}^0,\bm{x}_{D_2}^0,\alpha)$ denote the AS and BAS determined by two defenders $P_{D_1},P_{D_2}$ and one attacker $P_A$, respectively. Similarly, the AS is the set of points in $\mathbb{R}^n$ that $P_A$ can reach before both two defenders, which is formally stated below:
\begin{equation}\label{AR2}\begin{aligned}
\mathcal{R}_A^2(\bm{x}_A^0,\bm{x}_{D_1}^0,\bm{x}_{D_2}^0,\alpha)=&\{\bm{z}\in\mathbb{R}^n|\|\bm{z}-\bm{x}_A^0\|_2\\&< \alpha \|\bm{z}-\bm{x}_{D_i}^0\|_2,i=1,2\}.
\end{aligned}\end{equation}
Then, (\ref{AR2}) can also be equivalently rewritten as
\begin{equation*}\label{AR22}\begin{aligned}
\mathcal{R}_A^2(\bm{x}_A^0,\bm{x}_{D_1}^0,\bm{x}_{D_2}^0,\alpha)&=\{\bm{z}\in\mathbb{R}^n|\|\bm{z}-\theta_i\|_2<\delta_i,i=1,2\}.
\end{aligned}\end{equation*}
Hence, $\mathcal{R}_A^2(\bm{x}_A^0,\bm{x}_{D_1}^0,\bm{x}_{D_2}^0,\alpha)$ is the intersection set of two balls' interiors $\mathcal{R}_A^1(\bm{x}_A^0,\bm{x}_{D_1}^0,\alpha)$ and $\mathcal{R}_A^1(\bm{x}_A^0,\bm{x}_{D_2}^0,\alpha)$. Naturally, its boundary is ${\rm bas}^2(\bm{x}_A^0,\bm{x}_{D_1}^0,\bm{x}_{D_2}^0,\alpha)$. 

Unless needed for clarity, to simplify notations, hereinafter, we drop the initial conditions and speed ratio occurring in the expressions defined before. It is worth emphasizing that the geometric meanings of $\theta_i,\delta_i,\mathcal{R}_A^1,{\rm bas}^1,\mathcal{R}_A^2$ and ${\rm bas}^2$ will be frequently used. Next, we present two important lemmas.

\begin{lema}[Matrix property]\label{matrix}\rm
The square matrix $C_{ij}$ is positive semidefinite, and the three matrixes $A_{ij},B_{ij}$ and $C_{ij}$ satisfy
\begin{equation}\begin{aligned}\label{matrixequ}
C_{ij}A_{ij}=0,A_{ij}^\mathsf{T}C_{ij}=0,C_{ij}C_{ij}=\|A_{ij}\|_2^2C_{ij}\\
C_{ij}\bm{x}_{D_i,-n}^0=C_{ij}\bm{x}_{D_j,-n}^0=C_{ij}B_{ij}.
\end{aligned}\end{equation}

\emph{Proof:} From Section \ref{notations}, it can be easily observed that $C_{ij}$ is symmetric. Take any $\bm{x}$ in $\mathbb{R}^{n-1}$, we have
\begin{equation*}
\bm{x}^\mathsf{T}C_{ij}\bm{x}=\|A_{ij}\|_2^2\|\bm{x}\|_2^2-(A_{ij}^\mathsf{T}\bm{x})^2\ge0.
\end{equation*}
Thus, $C_{ij}$ is positive semidefinite. 

According to the definition, note that
\begin{equation*}\begin{aligned}\label{matrixequ1122}
C_{ij}A_{ij}&=(\|A_{ij}\|_2^2I_{n-1}-A_{ij}A_{ij}^\mathsf{T})A_{ij}=0.\\
\end{aligned}\end{equation*}
Thus, $A_{ij}^\mathsf{T}C_{ij}=0$ also holds. Furthermore, we have\begin{equation*}\begin{aligned}\label{matrixequ1132}
C_{ij}C_{ij}=(\|A_{ij}\|_2^2I_{n-1}-A_{ij}A_{ij}^\mathsf{T})C_{ij}=\|A_{ij}\|_2^2C_{ij}.
\end{aligned}\end{equation*}
By considering the definitions of $A_{ij}$ and $B_{ij}$ stated in Section \ref{notations}, it can be obtained that
\begin{equation*}\begin{aligned}\label{matrixequ11334}
C_{ij}A_{ij}=0\Rightarrow C_{ij}\bm{x}_{D_i,-n}^0=C_{ij}\bm{x}_{D_j,-n}^0=C_{ij}B_{ij}.
\end{aligned}\end{equation*}
Thus, we finish the proof.
\qed
\end{lema}

\begin{lema}[Optimal trajectories]\label{bound}\rm
Consider the system (\ref{dynamics}) satisfying Assumptions \ref{aspisolation}-\ref{aspratio}. If two defenders can guarantee to capture the attacker in $\Omega_{\rm play}$, namely, $\bm{x}_A^0\in\mathcal{W}_D^2$, $J_\mathcal{T}$ in (\ref{DDSpf}) is adopted by two teams, and if the attacker can assure its arrival in $\Omega_{\rm tar}$, namely, $\bm{x}_A^0\in\mathcal{W}_A^2$, $J_d$ in (\ref{ADSpf}) is adopted by two teams. Then, the optimal headings of $P_{D_1},P_{D_2}$ and $P_A$ are constant and the optimal trajectories are straight lines. 
\end{lema}
\emph{Proof:} The proof follows from the fact that three players have simple motion and the payoff function $J_\mathcal{T}$ in (\ref{DDSpf}) or $J_d$ in (\ref{ADSpf}) is of Meyer type.\qed

\subsection{One Defender Versus One Attacker}\label{subsectionb1v1}
We first present the barrier and winning subspaces for the case with one defender $P_{D_i}$ and one attacker $P_A$, which will provide key insights into the barrier construction for the two-defender scenario.

Suppose that $P_A$ initially lies at a position from which $P_A$ can reach $\mathcal{T}$ without being captured, and denote its optimal target point (OTP) in $\mathcal{T}$ by  $\bm{p}^*$ such that the payoff function $J_d$ in (\ref{ADSpf}) involving only one defender $P_{D_i}$ is maximized.

Note that the non-anticipative information structure implies that $P_{D_i}$ knows $P_A$'s current input. Since $P_{D_i}$ aims to minimize $J_d$ in (\ref{ADSpf}), the optimal strategy for $P_{D_i}$ is to move towards the same target point, i.e., $\bm{p}^*$, in $\mathcal{T}$ as $P_A$ does. Hence, by combining with \autoref{bound}, in one defender case, the payoff function $J_d$ involving only one defender $P_{D_i}$, is equivalent to the following function defined on the TH $\mathcal{T}$:
\begin{equation}\label{disfunc1v1}
F_i(\bm{p})=\|\bm{p}-\bm{x}_{D_i}^0\|_2-\frac{\|\bm{p}-\bm{x}_A^0\|_2}{\alpha},\bm{p}\in\mathcal{T}
\end{equation}
representing the distance between $P_{D_i}$ and $P_A$ exactly when $P_A$ reaches a point $\bm{p}$ in $\mathcal{T}$, as $P_{D_i}$ and $P_A$ move towards $\bm{p}$ with straight trajectories. If $P_{D_i}$ and $P_A$ both adopt their optimal strategies to minimize and maximize $F_i(\bm{p})$ respectively, then $\bm{p}^*$ must be an extreme point of $F_i(\bm{p})$, namely,
\begin{equation}\label{1v1FNC}
\frac{\partial F_i(\bm{p}^*)}{\partial\bm{p_{-n}}}=0\Rightarrow\frac{\bm{p}^*_{-n}-\bm{x}_{D_i,-n}^0}{\|\bm{p}^*-\bm{x}_{D_i}^0\|_2}=\frac{\bm{p}^*_{-n}-\bm{x}_{A,-n}^0}{\alpha\|\bm{p}^*-\bm{x}_A^0\|_2}
\end{equation}
by noting that $\bm{p},\bm{p}^*\in\mathcal{T}$ implies that $p_n=p_n^*=0$. We can conclude that if $\bm{p}^*$ is an OTP in $\mathcal{T}$, it must satisfy (\ref{1v1FNC}). Due to its highly frequent use, (\ref{1v1FNC}) is called the first-order necessary condition for an OTP, and for short, it is also called FNC.
\begin{lema}[Barrier and winning subspaces for one defender]\label{1v1barrierlema1less1}\rm
Suppose that Assumptions \ref{aspisolation}-\ref{aspratio} hold. If the system (\ref{dynamics}) has only one defender $P_{D_i}$, then the barrier $\mathcal{B}^1$ and two winning subspaces $\mathcal{W}_D^1$ and $\mathcal{W}_A^1$ are respectively given by\begin{equation}\begin{aligned}\label{barrier1v1less1}
&\mathcal{B}^1(\bm{x}_{D_i}^0,\alpha)=\{\bm{z}\in\mathbb{R}^n|Z=[\bm{z}^\mathsf{T},1]^\mathsf{T},\\
&\qquad\qquad\qquad\qquad\qquad\qquad Z^\mathsf{T}\Xi_iZ=0,z_n>0\}\\
&\mathcal{W}_D^1(\bm{x}_{D_i}^0,\alpha)=\{\bm{z}\in\mathbb{R}^n|Z=[\bm{z}^\mathsf{T},1]^\mathsf{T},\\
&\qquad\qquad\qquad\qquad\qquad\qquad Z^\mathsf{T}\Xi_iZ>0,z_n>0\}\\
&\mathcal{W}_A^1(\bm{x}_{D_i}^0,\alpha)=\{\bm{z}\in\mathbb{R}^n|Z=[\bm{z}^\mathsf{T},1]^\mathsf{T},\\
&\qquad\qquad\qquad\qquad\qquad\qquad Z^\mathsf{T}\Xi_iZ<0,z_n>0\}
\end{aligned}\end{equation}
and when $\bm{x}_A^0\in\mathcal{B}^1$, the OTP $\bm{p}^*$ in $\mathcal{T}$ for $P_{D_i}$ and $P_A$ is uniquely given by 
\begin{equation}\label{OTP1v1less1}
\bm{p}_{-n}^*=\theta_{i,-n},p_n^*=0.
\end{equation}
\end{lema}
\emph{Proof:} Consider $\bm{x}_A^0\in\mathcal{B}^1$ and assume that $\bm{p}^*$ is its OTP in $\mathcal{T}$ such that $J_d$ in (\ref{ADSpf}) is maximized, then FNC (\ref{1v1FNC}) holds when $P_{D_i}$ strives to minimize $J_d$. Furthermore, according to \autoref{bound}, $\bm{x}_A^0\in\mathcal{B}^1$ implies that \begin{equation}\label{barrierdist1v1less1}
\|\bm{p}^*-\bm{x}_A^0\|_2=\alpha\|\bm{p}^*-\bm{x}_{D_i}^0\|_2
\end{equation}
which reflects the fact that when $P_A$ initially lies at the barrier $\mathcal{B}^1$, $P_{D_i}$ and $P_A$ will reach the OTP $\bm{p}^*$ at the same time if their respective optimal strategies are adopted. In other words, if $P_A$ initially lies at the barrier $\mathcal{B}^1$, no player can win the game, that is, the capture and arrival occur at the same time. Thus, it follows from the FNC (\ref{1v1FNC}) and (\ref{barrierdist1v1less1}) that 
\begin{equation}\label{OTP1v1less222}
(1-\alpha^2)\bm{p}^*_{-n}=\bm{x}_{A,-n}^0-\alpha^2\bm{x}_{D_i,-n}^0.
\end{equation}
Also notice that $\bm{p}^*\in\mathcal{T}$ implies that $p_n^*=0$. Hence, when $\bm{x}_A^0\in\mathcal{B}^1$, the OTP $\bm{p}^*$ for $P_{D_i}$ and $P_A$ is uniquely given by (\ref{OTP1v1less1}), where $\theta_{i}$ is defined in (\ref{thetaanddelta}).

Since $p^*_n=0$, (\ref{barrierdist1v1less1}) can be simplified as
\begin{equation}\begin{aligned}\label{barrier1v1less1deriv22327}
&\|\bm{p}^*_{-n}-\bm{x}_{A,-n}^0\|_2^2+(x_{A,n}^0)^2\\
&\qquad\qquad\qquad=\alpha^2\|\bm{p}^*_{-n}-\bm{x}_{D_i,-n}^0\|_2^2+\alpha^2(x_{D_i,n}^0)^2.
\end{aligned}\end{equation}
Then, substituting $\bm{p}_{-n}^*$ given by (\ref{OTP1v1less222}) into (\ref{barrier1v1less1deriv22327}) leads to\begin{equation}
\begin{aligned}\label{barrier1v1less1deriv2}
&\|\frac{\bm{x}_{A,-n}^0-\alpha^2\bm{x}_{D_i,-n}^0}{1-\alpha^2}-\bm{x}_{A,-n}^0\|_2^2+(x_{A,n}^0)^2\\
&=\alpha^2\|\frac{\bm{x}_{A,-n}^0-\alpha^2\bm{x}_{D_i,-n}^0}{1-\alpha^2}-\bm{x}_{D_i,-n}^0\|_2^2+\alpha^2(x_{D_i,n}^0)^2\\
&\Rightarrow(x_{A,n}^0)^2-\frac{\alpha^2\|\bm{x}_{A,-n}^0-\bm{x}_{D_i,-n}^0\|_2^2}{1-\alpha^2}-\alpha^2(x_{D_i,n}^0)^2=0\\
&\Rightarrow(1/\alpha^2-1)(x_{A,n}^0)^2-\|\bm{x}_{A,-n}^0\|^2_2+2(\bm{x}_{D_i,-n}^0)^\mathsf{T}\bm{x}_{A,-n}^0\\
&\qquad\qquad\qquad\qquad\quad\ +\alpha^2(x_{D_i,n}^0)^2-\|\bm{x}_{D_i}^0\|_2^2=0
\end{aligned}\end{equation}
which characterizes the relationship between initial positions of $P_{D_i}$ and $P_A$ when $\bm{x}_A^0\in\mathcal{B}^1$. Thus, given $\bm{x}_{D_i}^0$, by taking all positions of $\bm{x}_{A,-n}^0$ in $\mathbb{R}^{n-1}$, $x_{A,n}^0$ can be explicitly computed from (\ref{barrier1v1less1deriv2}), and thus all positions of $P_A$ lying at $\mathcal{B}^1$ are found. Equivalently, these initial positions of $P_A$ form the barrier $\mathcal{B}^1$.

Define $Z=[(\bm{x}_A^0)^\mathsf{T},1]^\mathsf{T}$. By matrix multiplication, a compact formulation for the last equation in (\ref{barrier1v1less1deriv2}) can be achieved:\begin{equation*}\begin{aligned}\label{barrier1v1less1deriv2compact}
Z^\mathsf{T}\Xi_iZ=0
\end{aligned}\end{equation*}
where the matrix $\Xi_i$ is defined in (\ref{Ximatrix1122}). Note that \autoref{asprelaxed} and (\ref{new_tar}) imply that $x_{A,n}^0>0$ holds. Thus, the barrier $\mathcal{B}^1$ is given by (\ref{barrier1v1less1}). 

Notice that $1/\alpha^2-1>0$ for the last equation in (\ref{barrier1v1less1deriv2}). Since $\mathcal{W}_D^1$ and $\mathcal{W}_A^1$ are two subspaces of $\Omega_{\rm play}$ and separated by $\mathcal{B}^1$, and $\mathcal{W}_A^1$ is closer to $\mathcal{T}$ than $\mathcal{W}_D^1$, it follows from the last equation in (\ref{barrier1v1less1deriv2}) that these two winning subspaces satisfy the constraint stated in (\ref{barrier1v1less1}).  
\qed

\subsection{Two Defenders Versus One Attacker}\label{subsectionb2v1}
As will be shown below, $\mathcal{B}^2$ has two types. The first one is only dependent on one of two defenders, and the second one is associated with both two defenders. The conditions to distinguish them are as follows. For clarity, the defender which contributes to the barrier, is called \emph{active defender}.

\begin{lema}[Classification conditions]\rm\label{twotypeslema}
Suppose that Assumptions \ref{aspisolation}-\ref{aspratio} hold. 
\begin{itemize}
\item[a](Two active defenders). The barrier $\mathcal{B}^2$ is associated with both two defenders, if and only if\begin{equation}\begin{aligned}\label{equ1:onefromtwo}
&\bm{x}_{D_1,-n}^0\neq\bm{x}_{D_2,-n}^0, \text{ or }\\
&\bm{x}_{D_1,-n}^0=\bm{x}_{D_2,-n}^0,x_{D_1,n}^0=-x_{D_2,n}^0.
\end{aligned}\end{equation}
\item[b](One active defender). Otherwise, for $\bm{x}_{D_1,-n}^0=\bm{x}_{D_2,-n}^0$, if $|x_{D_1,n}^0|<|x_{D_2,n}^0|$, the barrier $\mathcal{B}^2$ only depends on $P_{D_1}$, i.e., $\mathcal{B}^2=\mathcal{B}^1(\bm{x}^0_{D_1},\alpha)$, and if $|x_{D_1,n}^0|>|x_{D_2,n}^0|$, the barrier $\mathcal{B}^2$ only depends on $P_{D_2}$, i.e., $\mathcal{B}^2=\mathcal{B}^1(\bm{x}^0_{D_2},\alpha)$.
\end{itemize}
\end{lema}
\emph{Proof:} Consider case (a) first. Assume that (\ref{equ1:onefromtwo}) holds. If $\bm{x}_{D_1,-n}^0\neq\bm{x}_{D_2,-n}^0$, namely, $A_{12}$ is a nonzero vector, it can be seen that $\mathcal{T}$'s three subspaces $\mathcal{N}_1,\mathcal{N}_{12}$ and $\mathcal{N}_2$ in (\ref{subspacesofTS}) are all nonempty. Thus, there must exist points in $\mathcal{T}$ for each defender that it can reach before the other defender. 

Let $\bm{p}$ be a point in $\mathcal{T}$ that $P_{D_1}$ can reach before $P_{D_2}$, i.e., $\bm{p}\in\mathcal{N}_1$. Thus, the attacker whose initial position lies at $\mathcal{B}^2$ with $\bm{p}$ as its OTP, could be captured assuringly only by $P_{D_1}$ while beyond the capability of $P_{D_2}$. Therefore, $\mathcal{B}^2$ depends on $P_{D_1}$. In the similar way, it can be obtained that $\mathcal{B}^2$ also depends on $P_{D_2}$. Thus, $\mathcal{B}^2$ depends on both two defenders.

If $\bm{x}_{D_1,-n}^0=\bm{x}_{D_2,-n}^0$ and $x_{D_1,n}^0=-x_{D_2,n}^0$, it can be seen that two defenders are symmetric with respect to $\mathcal{T}$, implying that for any point $\bm{p}$ in $\mathcal{T}$, two defenders can reach with the same time. Thus, $\mathcal{B}^2$ depends on both two defenders. Actually, the mirror property stated in \autoref{mirror} also reveals that two defenders symmetric with respect to $\mathcal{T}$ play the same role in barrier construction, as \autoref{mirrorrek} illustrates.

Conversely, assume that $\bm{x}_{D_1,-n}^0=\bm{x}_{D_2,-n}^0 $ and $x_{D_1,n}^0\neq-x_{D_2,n}^0$. If $|x_{D_1,n}^0|<|x_{D_2,n}^0|$, for any point $\bm{p}$ in $\mathcal{T}$, we have\begin{equation*}\begin{aligned}\label{equ1:onefromtwo1122}
\|\bm{x}_{D_1}^0-\bm{p}\|_2<\|\bm{x}_{D_2}^0-\bm{p}\|_2
\end{aligned}\end{equation*}
meaning that $P_{D_1}$ can reach any position in $\mathcal{T}$ before $P_{D_2}$, and this feature guarantees that the barrier $\mathcal{B}^2$ is determined by $P_{D_1}$ alone, that is, $\mathcal{B}^2=\mathcal{B}^1(\bm{x}^0_{D_1},\alpha)$. If $|x_{D_1,n}^0|>|x_{D_2,n}^0|$, analogously, it can be concluded that the barrier $\mathcal{B}^2$ is determined by $P_{D_2}$ alone, that is, $\mathcal{B}^2=\mathcal{B}^1(\bm{x}^0_{D_2},\alpha)$. Thus, the case (a) is proved.

According to the argument for case (a), the conclusion in case (b) is straightforward.\qed

Next, we construct the barrier and winning subspaces. As \autoref{twotypeslema} states, the ones for one active defender case can be obtained directly from Section \ref{subsectionb1v1}. Thus, our attention will be paid to the two active defender case where both two defenders make contributions to the construction of the barrier.

\begin{thom}[Barrier and winning subspaces for two active defenders]\label{thom2v1barrierless1}\rm
Consider the system (\ref{dynamics}) satisfying Assumptions \ref{aspisolation}-\ref{aspratio}. Suppose that (\ref{equ1:onefromtwo}) is true. If $\bm{x}_{D_1,-n}^0\neq\bm{x}_{D_2,-n}^0$, the barrier $\mathcal{B}^2$ is given by $\bigcup_{i=1}^3\mathcal{B}_i^2$, which consists of three parts:\begin{equation}\begin{aligned}\label{equ1:twod:zero:less1}
&\mathcal{B}^2_i(\bm{x}^0_{D_1},\bm{x}^0_{D_2},\alpha)=\{\bm{z}\in\mathbb{R}^n|Z=[\bm{z}^\mathsf{T},1]^\mathsf{T},\\
&\qquad\qquad Z^\mathsf{T}\Xi_iZ=0,z_n>0,\bm{z}\in \mathcal{A}_{i}\}(i=1,2)\\
&\mathcal{B}^2_3(\bm{x}^0_{D_1},\bm{x}^0_{D_2},\alpha)=\{\bm{z}\in\mathbb{R}^n|Z=[\bm{z}^\mathsf{T},1]^\mathsf{T},\\
&\qquad\qquad\qquad\quad Z^\mathsf{T}\Xi_{12}Z=0,z_n>0,\bm{z}\in \mathcal{A}_{12}\}\\
\end{aligned}\end{equation}
where the constraints for the first $n-1$ dimensions of $\bm{z}$ are
\begin{equation}\begin{aligned}\label{barrierpara2v1}
\mathcal{A}_i&=\{\bm{z}\in\mathbb{R}^n|A_{ij}^\mathsf{T}\bm{z}_{-n}-\alpha^2A_{ij}^\mathsf{T}\bm{x}_{D_i,-n}^0\\
&\qquad\quad\ >(1-\alpha^2)w_{ij}\}(i,j=1,2,i\neq j)\\
\mathcal{A}_{12}&=\{\bm{z}\in\mathbb{R}^n|\bm{z}\in\mathbb{R}^{n}\setminus(\mathcal{A}_1\cup \mathcal{A}_2)\}.\\
\end{aligned}\end{equation}
The winning subspace $\mathcal{W}_D^2$ is the subspace given by (\ref{equ1:twod:zero:less1}) with the expression $=0$ replaced by $>0$, and $\mathcal{W}_A^2$ corresponds to the case by replacing  $=0$ in (\ref{equ1:twod:zero:less1}) with $<0$. If $\bm{x}_{D_1,-n}^0=\bm{x}_{D_2,-n}^0$ and $x_{D_1,n}^0=-x_{D_2,n}^0$, the barrier $\mathcal{B}^2$, two winning subspaces $\mathcal{W}_D^2$ and $\mathcal{W}_A^2$ are given by
\begin{equation}\begin{aligned}\label{barrierpara2v1223344}
\mathcal{B}^2(\bm{x}^0_{D_1},\bm{x}^0_{D_2},\alpha)&=\mathcal{B}^1(\bm{x}_{D_1}^0,\alpha)=\mathcal{B}^1(\bm{x}_{D_2}^0,\alpha)\\
\mathcal{W}_D^2(\bm{x}^0_{D_1},\bm{x}^0_{D_2},\alpha)&=\mathcal{W}_D^1(\bm{x}_{D_1}^0,\alpha)=\mathcal{W}_D^1(\bm{x}_{D_2}^0,\alpha)\\
\mathcal{W}_A^2(\bm{x}^0_{D_1},\bm{x}^0_{D_2},\alpha)&=\mathcal{W}_A^1(\bm{x}_{D_1}^0,\alpha)=\mathcal{W}_A^1(\bm{x}_{D_2}^0,\alpha).
\end{aligned}\end{equation}
\end{thom}
\emph{Proof:} Since (\ref{equ1:onefromtwo}) holds, it follows from \autoref{twotypeslema} that $\mathcal{B}^2$ depends on both two defenders. We first consider the case $\bm{x}_{D_1,-n}^0\neq\bm{x}_{D_2,-n}^0$. 

Then, as the proof of \autoref{twotypeslema} shows, there exist points in $\mathcal{T}$ for each defender that it can reach before the other defender. Hence, three associated subspaces $\mathcal{N}_1,\mathcal{N}_{12}$ and $\mathcal{N}_2$ in (\ref{subspacesofTS}) are all nonempty. Assume $\bm{x}_A^0\in\mathcal{B}^2$ and denote its OTP in $\mathcal{T}$ by $\bm{p}^*$ such that the payoff function $J_d$ in (\ref{ADSpf}) is maxmized. Define $Z=[(\bm{x}_A^0)^\mathsf{T},1]^\mathsf{T}$.

If $\bm{p}^*$ lies in the subspace $\mathcal{N}_1$, then $P_{D_1}$ can reach $\bm{p}^*$ before $P_{D_2}$. Thus, in this case, the barrier only depends on $P_{D_1}$, and according to \autoref{1v1barrierlema1less1}, the barrier can be constructed by (\ref{barrier1v1less1}) with $i=1$ as follows:
\begin{equation*}\begin{aligned}\label{barrier1v1less1deriv2compact22}
Z^\mathsf{T}\Xi_1Z=0,x_{A,n}^0>0
\end{aligned}\end{equation*}
and by (\ref{OTP1v1less1}), the OTP $\bm{p}^*$ for $P_{D_1}$ and $P_A$ is given by\begin{equation}\label{pd11v1less1}
\bm{p}_{-n}^*=\theta_{1,-n},p_n^*=0.
\end{equation}
Since $\bm{p}^*\in\mathcal{N}_1$, (\ref{subspacesofTS}) and (\ref{pd11v1less1}) lead to
\begin{equation*}\begin{aligned}\label{pd11v1less12}
A_{12}^\mathsf{T}\theta_{1,-n}>w_{12}\Rightarrow A_{12}^\mathsf{T}(\bm{x}_{A,-n}^0-\alpha^2\bm{x}_{D_1,-n}^0)>(1-\alpha^2)w_{12}
\end{aligned}\end{equation*}
representing the constraint that $\bm{x}_{A,-n}^0$ satisfies. This constraint can be equivalently reformulated as $\bm{x}_A^0\in\mathcal{A}_1$, where $\mathcal{A}_1$ is defined in (\ref{barrierpara2v1}). Therefore, the part of $\mathcal{B}^2$ which only depends on $P_{D_1}$, denoted by $\mathcal{B}^2_1$, is obtained as (\ref{equ1:twod:zero:less1}) shows.

Analogously, if $\bm{p}^*$ lies in the subspace $\mathcal{N}_2$, we can obtain that the part of $\mathcal{B}^2$ which only depends on $P_{D_2}$, denoted by $\mathcal{B}^2_2$, is also as (\ref{equ1:twod:zero:less1}) describes.

Now, turn to the remaining part $\mathcal{B}^2_3$ of $\mathcal{B}^2$ when $\bm{p}^*$ lies in the subspace $\mathcal{N}_{12}$, namely, two defenders capture $P_A$ at the same time and thus limit the OTP $\bm{p}^*$ in $\mathcal{N}_{12}$. Therefore, the OTP $\bm{p}^*$ for three players is an extreme point of the function $F_1(\bm{p})$ in (\ref{disfunc1v1}) with the constraint $\bm{p}\in\mathcal{N}_{12}$.

Note that $\mathcal{N}_{12}$ is given by (\ref{subspacesofTS}), and thus define the associated Hamiltonian function
\begin{equation*}\label{hamilton}
H_1(\bm{p},\lambda)=F_1(\bm{p})+\lambda(A_{12}^\mathsf{T}\bm{p}_{-n}-w_{12}),p_n=0
\end{equation*}
where $\lambda\in\mathbb{R}$ is the Lagrangian multiplier. Therefore, the OTP $\bm{p}^*$ meets the optimality conditions of $H_1$ as follows:
\begin{subequations}\label{barrier1v1less1optcondition}\begin{align}
\frac{\partial H_1}{\partial\bm{p}_{-n}}&=\frac{\bm{p}^*_{-n}-\bm{x}_{D_1,-n}^0}{\|\bm{p}^*-\bm{x}_{D_1}^0\|_2}-\frac{\bm{p}^*_{-n}-\bm{x}_{A,-n}^0}{\alpha\|\bm{p}^*-\bm{x}_A^0\|_2}+A_{12}\lambda =0\label{barrier1v1less1optcondition1}\\
\frac{\partial H_1}{\partial\lambda}&=A_{12}^\mathsf{T}\bm{p}_{-n}^*-w_{12}=0.\label{barrier1v1less1optcondition2}
\end{align}\end{subequations}

Since $p_n^*=0$, in what follows, we focus on $\bm{p}_{-n}^*$. Note that $\bm{x}_A^0\in\mathcal{B}^2$ implies that $P_A$ reaches the OTP $\bm{p}^*$ exactly when captured, so we have (\ref{barrierdist1v1less1}) with $i=1$, and substituting it into (\ref{barrier1v1less1optcondition1}) yields
\begin{equation}\begin{aligned}\label{barrier1v1less1123}
&\frac{\bm{x}_{A,-n}^0-\alpha^2\bm{x}_{D_1,-n}^0-(1-\alpha^2)\bm{p}_{-n}^*}{\alpha^2\|\bm{p}^*-\bm{x}_{D_1}^0\|_2}+A_{12}\lambda =0\\
&\Rightarrow\frac{(1-\alpha^2)(\theta_{1,-n}-\bm{p}_{-n}^*)}{\alpha^2\|\bm{p}^*-\bm{x}_{D_1}^0\|_2}+A_{12}\lambda =0.
\end{aligned}\end{equation}

Since $\bm{x}_{D_1,-n}^0\neq\bm{x}_{D_2,-n}^0$, then $\|A_{12}\|_2^2$ is positive. For (\ref{barrier1v1less1123}), by multiplying $A_{12}^\mathsf{T}$ from the left, we can obtain
\begin{equation}\label{lambdavalue1v1}
\lambda=-\frac{(1-\alpha^2)A_{12}^\mathsf{T}(\theta_{1,-n}-\bm{p}_{-n}^*)}{\alpha^2\|A_{12}\|_2^2\|\bm{p}^*-\bm{x}_{D_1}^0\|_2}
\end{equation}
and then substituting (\ref{lambdavalue1v1}) into (\ref{barrier1v1less1123}) and combining (\ref{barrier1v1less1optcondition2}) lead to $\bm{p}_{-n}^*$ explicitly given by
\begin{equation}\begin{aligned}\label{OTP2vs1B2explicitly}
&\bm{p}^*_{-n}=\theta_{1,-n}-\frac{A_{12}A_{12}^\mathsf{T}(\theta_{1,-n}-\bm{p}_{-n}^*)}{\|A_{12}\|_2^2}\\
&=\theta_{1,-n}-\frac{A_{12}A_{12}^\mathsf{T}\theta_{1,-n}-A_{12}w_{12}}{\|A_{12}\|_2^2}=\frac{C_{12}\theta_{1,-n}+A_{12}w_{12}}{\|A_{12}\|_2^2}.
\end{aligned}\end{equation}
Thus, we obtain the OTP $\bm{p}^*$ for three players when $\bm{x}_A^0\in\mathcal{B}_3^2$.

Next, the analytical description of $\mathcal{B}^2_3$ will be investigated. According to the property of simultaneous arriving (\ref{barrierdist1v1less1}) for $i=1$ and $p_n^*=0$, we have 
\begin{equation}\begin{aligned}\label{less1distequsquare}
&\|\bm{p}^*-\bm{x}_{A}^0\|_2^2=\alpha^2\|\bm{p}^*-\bm{x}_{D_1}^0\|_2^2\Rightarrow\|\bm{p}_{-n}^*\|_2^2\\
&-2\theta_{1,-n}^\mathsf{T}\bm{p}_{-n}^*+(\|\bm{x}_A^0\|_2^2-\alpha^2\|\bm{x}_{D_1}^0\|^2_2)/(1-\alpha^2)=0.
\end{aligned}\end{equation}
By (\ref{OTP2vs1B2explicitly}) and (\ref{matrixequ}), the two terms $\|\bm{p}_{-n}^*\|_2^2$ and $\theta_{1,-n}^\mathsf{T}\bm{p}_{-n}^*$ in (\ref{less1distequsquare}) can be computed as follows:
\begin{equation}\begin{aligned}\label{pninternal1}
\|\bm{p}_{-n}^*\|_2^2&=\frac{\theta_{1,-n}^\mathsf{T}C_{12}\theta_{1,-n}+w_{12}^2}{\|A_{12}\|_2^2}\\
\theta_{1,-n}^\mathsf{T}\bm{p}_{-n}^*&=\frac{\theta_{1,-n}^\mathsf{T}C_{12}\theta_{1,-n}+\theta_{1,-n}^\mathsf{T}A_{12}w_{12}}{\|A_{12}\|_2^2}.
\end{aligned}\end{equation}
Therefore, by (\ref{pninternal1}), (\ref{less1distequsquare}) can be simplified to
\begin{equation}\begin{aligned}\label{pninternal1simplifiedd2244}
&\frac{-\theta_{1,-n}^\mathsf{T}C_{12}\theta_{1,-n}+w_{12}^2-2\theta_{1,-n}^\mathsf{T}A_{12}w_{12}}{\|A_{12}\|_2^2}\\
&\qquad\qquad\qquad\qquad\qquad+\frac{\|\bm{x}_A^0\|_2^2-\alpha^2\|\bm{x}_{D_1}^0\|^2_2}{1-\alpha^2}=0.
\end{aligned}\end{equation}

Furthermore, considering the expression of $\theta_{1,-n}$ in (\ref{thetaanddelta}) and employing the property (\ref{matrixequ}), we can obtain two equalities
\begin{equation}\begin{aligned}\label{pninternal1simplifiedd33}
&\theta_{1,-n}^\mathsf{T}C_{12}\theta_{1,-n}=\frac{(\bm{x}_{A,-n}^0)^\mathsf{T}C_{12}\bm{x}_{A,-n}^0}{(1-\alpha^2)^2}\\
&\quad+\frac{-2\alpha^2(\bm{x}_{A,-n}^0)^\mathsf{T}C_{12}\bm{x}_{D_1,-n}^0+\alpha^4(\bm{x}_{D_1,-n}^0)^\mathsf{T}C_{12}\bm{x}_{D_1,-n}^0}{(1-\alpha^2)^2}\\
&=\frac{(\bm{x}_{A,-n}^0)^\mathsf{T}C_{12}\bm{x}_{A,-n}^0}{(1-\alpha^2)^2}\\
&\quad+\frac{-2\alpha^2(\bm{x}_{A,-n}^0)^\mathsf{T}C_{12}B_{12}+\alpha^4B^{\mathsf{T}}_{12}C_{12}B_{12}}{(1-\alpha^2)^2}
\end{aligned}\end{equation}
and 
\begin{equation}\begin{aligned}\label{pninternal1simplifiedd334}
\theta_{1,-n}^\mathsf{T}A_{12}=\frac{(\bm{x}_{A,-n}^0)^\mathsf{T}A_{12}-\alpha^2(\bm{x}_{D_1,-n}^0)^\mathsf{T}A_{12}}{1-\alpha^2}.
\end{aligned}\end{equation}
Therefore, by substituting these two equalities (\ref{pninternal1simplifiedd33}) and (\ref{pninternal1simplifiedd334}), the equation (\ref{pninternal1simplifiedd2244}) can be equivalently rewritten as
\begin{equation}\begin{aligned}\label{barrier2v1less1ori22}
\zeta_{12,1}(x_{A,n}^0)^2=(\bm{x}_{A,-n}^0)^\mathsf{T}\zeta_{12,2}\bm{x}_{A,-n}^0+2\zeta_{12,3}^\mathsf{T}\bm{x}_{A,-n}^0+\zeta_{12,4}
\end{aligned}\end{equation}
where $\zeta_{12,i}(i=1,2,3,4)$ is defined in (\ref{zetadefinition}) which only depends on two defenders' initial positions and the speed ratio. It is worth noting that in computing the term $\zeta_{12,4}$ for (\ref{barrier2v1less1ori22}), the following equality is used:\begin{equation*}\begin{aligned}\label{barrier2v1les12345}
A_{12}\|\bm{x}_{D_1}^0\|_2^2-2\bm{x}_{D_1,-n}^0w_{12}\qquad\qquad\qquad\qquad\qquad\qquad\qquad&\\
=(\bm{x}_{D_1,-n}^0-\bm{x}_{D_2,-n}^0)\|\bm{x}_{D_1}^0\|_2^2-\bm{x}_{D_1,-n}^0(\|\bm{x}_{D_1}^0\|_2^2-\|\bm{x}_{D_2}^0\|_2^2)&\\
=\|\bm{x}_{D_2}^0\|_2^2\bm{x}_{D_1,-n}^0-\|\bm{x}_{D_1}^0\|_2^2\bm{x}_{D_2,-n}^0.\qquad\qquad\qquad\qquad\qquad&
\end{aligned}\end{equation*}
In a compact form, (\ref{barrier2v1less1ori22}) can also be rewritten as
\begin{equation}\begin{aligned}\label{newbarrier2v1213}
Z^\mathsf{T}\Xi_{12}Z=0
\end{aligned}\end{equation}
where the matrix $\Xi_{12}$ is defined in (\ref{Ximatrix1122}). Therefore, (\ref{newbarrier2v1213}) is the condition that $\bm{x}_A^0$ should satisfy when $P_A$ lies at $\mathcal{B}^2_3$.

Denote by $\mathcal{A}_{12}$ the constraint for $\bm{x}_{A,-n}^0$ when $\bm{p}^*\in\mathcal{N}_{12}$. As (\ref{subspacesofTS1}) and (\ref{subspacesofTS}) shows, $\mathcal{N}_{12}$ is the complementary of the union space of $\mathcal{N}_1$ and $\mathcal{N}_2$ in $\mathcal{T}$, and thus so is $\mathcal{A}_{12}$ with respect to $\mathcal{A}_1$ and $\mathcal{A}_2$ in $\mathbb{R}^{n}$. Hence, $\mathcal{B}_3^2$ is obtained analytically as (\ref{equ1:twod:zero:less1}) and (\ref{barrierpara2v1}) show. The construction of the barrier $\mathcal{B}^2$ is finished.

As (\ref{barrier1v1less1deriv2}) and (\ref{barrier2v1less1ori22}) show, it can be noted that $(1/\alpha^2-1)>0$ in (\ref{barrier1v1less1deriv2}) and $\zeta_{12,1}>0$ in (\ref{barrier2v1less1ori22}), and actually $\mathcal{B}^2$ is constructed by taking all positions in the first $n-1$ dimensional subspace and then computing their values in the $n$-th dimensional subspace based on (\ref{barrier1v1less1deriv2}) or (\ref{barrier2v1less1ori22}) classified by (\ref{barrierpara2v1}). Thus, for two winning subspaces $\mathcal{W}_D^2$ and $\mathcal{W}_A^2$, since $\mathcal{W}_D^2$ is farther from $\mathcal{T}$ than $\mathcal{W}_A^2$ and they both belong to the subspace $\Omega_{\rm play}$, the analytical descriptions of them are straightforward as this theorem states. 

Now, we consider the case $\bm{x}_{D_1,-n}^0=\bm{x}_{D_2,-n}^0$ and $x_{D_1,n}^0=-x_{D_2,n}^0$. It can be verified that $A_{12}$ is a zero vector and $w_{12}=0$. Then, (\ref{subspacesofTS}) shows that $\mathcal{N}_1=\mathcal{N}_2=\emptyset$ and $\mathcal{N}_{12}=\mathcal{T}$, implying that for any point in $\mathcal{T}$, two defenders can reach with the same time. Hence, $\mathcal{B}^2=\mathcal{B}^1(\bm{x}_{D_1}^0,\alpha)=\mathcal{B}^1(\bm{x}_{D_2}^0,\alpha)$, as (\ref{barrierpara2v1223344}) shows. The winning subspaces $\mathcal{W}_D^2$ and $\mathcal{W}_A^2$ are straightforward. Thus, we finish the proof.\qed
\begin{rek}\label{barrierOTP2v1less1}\rm
As the proof of \autoref{thom2v1barrierless1} indicates, if $\bm{x}_A^0\in\mathcal{B}_i^2(i=1,2)$, the OTP for $P_{D_i}$ and $P_A$ is unique and given by (\ref{OTP1v1less1}), and $P_{D_j}(j\neq i)$ can adopt any strategy in $\mathcal{U}$. If $\bm{x}_A^0\in\mathcal{B}_3^2$, the OTP $\bm{p}^*=[\bm{p}_{-n}^{*\mathsf{T}},0]^\mathsf{T}$ for three players is uniquely given by (\ref{OTP2vs1B2explicitly}) which can be also rewritten as follows:\begin{equation*}\begin{aligned}\label{OTP2v1barrierless1re}
\bm{p}^*_{-n}&=\frac{C_{12}\theta_{1,-n}+A_{12}w_{12}}{\|A_{12}\|_2^2}\\
&=\frac{C_{12}\bm{x}_{A,-n}^0-\alpha^2C_{12}\bm{x}_{D_1,-n}^0+(1-\alpha^2)A_{12}w_{12}}{(1-\alpha^2)\|A_{12}\|_2^2}\\
&=\frac{C_{12}\bm{x}_{A,-n}^0-\alpha^2C_{12}B_{12}+(1-\alpha^2)A_{12}w_{12}}{(1-\alpha^2)\|A_{12}\|_2^2}
\end{aligned}\end{equation*}
which reflects the fact that two defenders play the equal role in capturing $P_A$ (i.e., determining the OTP $\bm{p}^*$) in this condition.
\end{rek}

\section{Optimal Strategies in the DWS}\label{Optimal Strategies}
In this section, Problem \ref{degreeDDS} will be investigated to provide optimal strategies for three players when their initial positions lie in the DWS.

Since two defenders can guarantee to capture $P_A$ in $\Omega_{\rm play}$ when the latter lies in the DWS $\mathcal{W}_D^2$, $J_\mathcal{T}$ in (\ref{DDSpf}) is considered by two teams. Denote the capture point by $\bm{p}^*\in\mathbb{R}^n$, which is also the OTP for $P_A$ such that $J_\mathcal{T}$ is minimized. Thus, $J_\mathcal{T}$ can be rewritten as the distance between $\bm{p}^*$ and $\mathcal{T}$, namely
\begin{equation*}
J_{\mathcal{T}}(\bm{d},\bm{a};\bm{x}_{D_1}^0,\bm{x}_{D_2}^0,\bm{x}_A^0)=\min_{\bm{p}\in\mathcal{T}}\|\bm{p}-\bm{p}^*\|_2.
\end{equation*}

For convenience, define one map $\phi:\mathbb{R}^n\times\mathbb{R}^n\rightarrow\mathbb{R}^n$ satisfying $\phi(\bm{x},\bm{y})=\frac{\bm{x}-\bm{y}}{\|\bm{x}-\bm{y}\|_2}$ with $\bm{x}\neq\bm{y}$. According to the definition and geometric meaning of the BAS ${\rm bas}^2$ described in Section \ref{sectionthreeA}, the OTP for $P_A$ lies on ${\rm bas}^2$, i.e., $\bm{p}^*\in{\rm bas}^2$. Then, it follows from \autoref{bound} that the optimal strategy for $P_A$ is to travel towards $\bm{p}^*$ directly. Since $P_A$ aims to minimize $J_\mathcal{T}$, the following lemma can be obtained directly.
\begin{lema}\label{DDROptStra}\rm
Consider the system (\ref{dynamics}) satisfying Assumptions \ref{aspisolation}-\ref{aspratio}. If $\bm{x}_A^0\in\mathcal{W}_D^2$ and the payoff function $J_\mathcal{T}$ in (\ref{DDSpf}) is considered, then the capture point $\bm{p}^*$ is the closest point to $\mathcal{T}$ on ${\rm bas}^2$, and the optimal strategy for $P_A$ is $\bm{a}^*=\phi(\bm{p}^*,\bm{x}_A^0)$.
\end{lema}
Next, how to locate the capture point $\bm{p}^*$, namely, the OTP, is discussed. We call the defender which captures the attacker under all players' optimal plays, as \emph{effective defender}.
\begin{thom}[Optimal strategies for DWS]\label{thomDDS2v1less1}\rm
Consider the system (\ref{dynamics}) satisfying Assumptions \ref{aspisolation}-\ref{aspratio}. If $\bm{x}_A^0\in\mathcal{W}_D^2$ and the payoff function $J_\mathcal{T}$ in (\ref{DDSpf}) is considered, then the saddle-point equilibrium $(\bm{d}^*,\bm{a}^*)$ of (\ref{DDSvf}) is as follows:
\begin{itemize}
\item[a](One effective defender). If there exists an $i\in\{1,2\}$ such that $\theta_i-\delta_i\bm{e}_n\in\mathcal{R}_A^1(\bm{x}_A^0,\bm{x}_{D_j}^0,\alpha)$ holds for $j\neq i$, then $\bm{d}_j^*$ can take any strategy in $\mathcal{U}$, and $(\bm{d}_i^*,\bm{a}^*)$ is uniquely given by
\begin{equation*}\label{DDSless1oneD}
\bm{d}_i^*=\phi(\bm{p}^*,\bm{x}_{D_i}^0),\bm{a}^*=\phi(\bm{p}^*,\bm{x}_A^0)
\end{equation*}
where $\bm{p}^*$ is the OTP and given by $\theta_i-\delta_i\bm{e}_n$.
\item[b](Two effective defenders). Otherwise, both two defenders make contributions to (\ref{DDSvf}), and $(\bm{d}^*,\bm{a}^*)$ is given by
\begin{equation*}\label{DDSless1twoD}
\bm{d}_i^*=\phi(\bm{p}^*,\bm{x}_{D_i}^0),\bm{a}^*=\phi(\bm{p}^*,\bm{x}_A^0),i=1,2
\end{equation*}
where the OTP $\bm{p}^*$ has two cases: If $m_{12}=x_{D_1,n}^0-x_{D_2,n}^0\neq0$, without loss of generality, assume $m_{12}>0$, and then \begin{equation*}\begin{aligned}\label{DDSless1twoDOTP1}
\bm{p}^*_{-n}&=\sqrt{\frac{R_2^\mathsf{T}R_1^{-1}R_2-r_3}{A_{12}^\mathsf{T}R_1^{-1}A_{12}}}R_1^{-1}A_{12}-R_1^{-1}R_2\\
p^*_n&=\frac{w_{12}-A_{12}^\mathsf{T}\bm{p}^*_{-n}}{m_{12}}
\end{aligned}\end{equation*}and if $m_{12}=0$, then
\begin{equation}\begin{aligned}\label{DDSless1twoDOTP2}
\bm{p}^*_{-n}&=\frac{A_{12}w_{12}+C_{12}\theta_{1,-n}}{\|A_{12}\|_2^2}\\
p^*_n&=\theta_{1,n}-\sqrt{\delta_1^2-\|\bm{p}^*_{-n}-\theta_{1,-n}\|_2^2}.
\end{aligned}\end{equation}
\end{itemize}
\end{thom}
\emph{Proof:} Part a: As (\ref{ARtransfer112}) shows, the BAS ${\rm bas}^1$ determined by $P_{D_i}$ and $P_A$ is a sphere of radius $\delta_i$ centered at $\theta_i$:
\begin{equation}\label{BAS1v1less1new}
{\rm bas}^1(\bm{x}_A^0,\bm{x}_{D_i}^0,\alpha)=\{\bm{z}\in\mathbb{R}^n|\|\bm{z}-\theta_i\|_2=\delta_i\}
\end{equation}
showing that the unique closest point to $\mathcal{T}$ on ${\rm bas}^1$, is $\theta_i-\delta_i\bm{e}_n$, where $\bm{e}_n\in\mathbb{R}^n$ is the vector of its $n$-th element equal to $1$ and the other elements equal to $0$. 

Assume that there exists an $i\in\{1,2\}$ such that $\theta_i-\delta_i\bm{e}_n\in\mathcal{R}_A^1(\bm{x}_A^0,\bm{x}_{D_j}^0,\alpha)(j\neq i)$ holds. As stated in Section \ref{sectionthreeA}, $\mathcal{R}_A^2$ is the intersection set of two balls' interiors $\mathcal{R}_A^1(\bm{x}_A^0,\bm{x}_{D_1}^0,\alpha)$ and $\mathcal{R}_A^1(\bm{x}_A^0,\bm{x}_{D_2}^0,\alpha)$, and ${\rm bas}^2$ is the boundary of $\mathcal{R}_A^2$. Thus, it can be claimed that $\theta_i-\delta_i\bm{e}_n$ is the unique closest point to $\mathcal{T}$ on ${\rm bas}^2$. It follows from \autoref{DDROptStra} that the capture point $\bm{p}^*$, i.e., the OTP, for $P_A$ is 
\begin{equation*}\label{DWScapturedpoint11}
\bm{p}^*=\theta_i-\delta_i\bm{e}_n.
\end{equation*}

Since $\theta_i-\delta_i\bm{e}_n\in\mathcal{R}_A^1(\bm{x}_A^0,\bm{x}_{D_j}^0,\alpha)$, $P_A$ can reach $\theta_i-\delta_i\bm{e}_n$ before $P_{D_j}$. Thus, $P_{D_j}$ has no contribution to the capture of $P_A$, implying that any strategy in $\mathcal{U}$ can be chosen for $P_{D_j}$. It can be seen that $P_{D_i}$ is the unique effective defender, and its unique OTP such that $J_\mathcal{T}$ is maximized, is also $\theta_i-\delta_i\bm{e}_n$.

Part b: If $\theta_1-\delta_1\bm{e}_n\notin\mathcal{R}_A^1(\bm{x}_A^0,\bm{x}_{D_2}^0,\alpha)$ and $\theta_2-\delta_2\bm{e}_n\notin\mathcal{R}_A^1(\bm{x}_A^0,\bm{x}_{D_1}^0,\alpha)$ both hold, we can state that both two defenders have effect on (\ref{DDSvf}) and the OTP $\bm{p}^*$ must belong to the intersection space of ${\rm bas}^1(\bm{x}_A^0,\bm{x}_{D_1}^0,\alpha)$ and ${\rm bas}^1(\bm{x}_A^0,\bm{x}_{D_2}^0,\alpha)$. Since the goal of $P_A$ is to minimize its final distance to $\mathcal{T}$, it can be concluded that the OTP $\bm{p}^*$ is the solution of the minimization problem
\begin{equation}\begin{aligned}\label{oriminimizationless1}
\min_{\bm{z}\in\mathbb{R}^n}z_n,\text{ s.t. }\|\bm{z}-\theta_1\|_2=\delta_1,\|\bm{z}-\theta_2\|_2=\delta_2.
\end{aligned}\end{equation}

First, we prove that $A_{12}$ is a nonzero vector in this case. If $A_{12}$ is a zero vector, i.e., $\bm{x}_{D_1,-n}^0=\bm{x}_{D_2,-n}^0$, (\ref{thetaanddelta}) implies that $\theta_{1,-n}=\theta_{2,-n}$. Thus, the vector $\theta_1-\theta_2$ is perpendicular to $\mathcal{T}$. Note that $\theta_1$ and $\theta_2$ are the centers of two spheres ${\rm bas}^1(\bm{x}_A^0,\bm{x}_{D_1}^0,\alpha)$ and ${\rm bas}^1(\bm{x}_A^0,\bm{x}_{D_2}^0,\alpha)$ given by (\ref{ARtransfer112}) . Thus, $\theta_1-\delta_1\bm{e}_n\in\mathcal{R}_A^1(\bm{x}_A^0,\bm{x}_{D_2}^0,\alpha)$ or $\theta_2-\delta_2\bm{e}_n\in\mathcal{R}_A^1(\bm{x}_A^0,\bm{x}_{D_1}^0,\alpha)$ holds. Even if $\theta_1-\delta_1\bm{e}_n=\theta_2-\delta_2\bm{e}_n$, it can be taken as the limiting case of Part a. Thus, $A_{12}$ is a nonzero vector.

Next, we transform the problem (\ref{oriminimizationless1}) into a minimization problem with only one equality constraint. First, by (\ref{AR1}) and (\ref{BAS1v1less1new}), the two constraints in (\ref{oriminimizationless1}) can also be rewritten as
\begin{equation}\begin{aligned}\label{intersectionbar123}
\|\bm{z}-\bm{x}_A^0\|_2=\alpha\|\bm{z}-\bm{x}_{D_1}^0\|_2,\|\bm{z}-\bm{x}_A^0\|_2=\alpha\|\bm{z}-\bm{x}_{D_2}^0\|_2
\end{aligned}\end{equation}
and thus the difference for (\ref{intersectionbar123}) yields
\begin{equation}\begin{aligned}\label{intersectionbar}
&\|\bm{z}-\bm{x}_{D_1}^0\|_2=\|\bm{z}-\bm{x}_{D_2}^0\|_2\\
&\Rightarrow2(\bm{x}_{D_1}^0-\bm{x}_{D_2}^0)^\mathsf{T}\bm{z}+\|\bm{x}_{D_2}^0\|_2^2-\|\bm{x}_{D_1}^0\|_2^2=0\\
&\Rightarrow m_{12}z_n+A_{12}^\mathsf{T}\bm{z}_{-n}-w_{12}=0
\end{aligned}\end{equation}
where $m_{12}=x_{D_1,n}^0-x_{D_2,n}^0$ is defined in Section \ref{notations}. There are two cases depending on whether $m_{12}=0$ holds.

Case 1: If $m_{12}\neq0$, (\ref{intersectionbar}) can be rewritten as follows:
\begin{equation}\label{znfromzminusn}
z_n=\frac{w_{12}-A_{12}^\mathsf{T}\bm{z}_{-n}}{m_{12}}
\end{equation}
and then for the BAS (\ref{BAS1v1less1new}) with $i=1$, we have
\begin{equation}\begin{aligned}\label{zminusneuqtionless1}
&(z_n-\theta_{1,n})^2+\|\bm{z}_{-n}-\theta_{1,-n}\|_2^2=\delta_1^2\\
&\Rightarrow\Big(\frac{w_{12}-A_{12}^\mathsf{T}\bm{z}_{-n}}{m_{12}}-\theta_{1,n}\Big)^2+\|\bm{z}_{-n}-\theta_{1,-n}\|_2^2=\delta_1^2\\
&\Rightarrow\bm{z}_{-n}^\mathsf{T}R_1\bm{z}_{-n}+2R_2^\mathsf{T}\bm{z}_{-n}+r_3=0
\end{aligned}\end{equation}
where $R_1,R_2$ and $r_3$ depend on two defenders' initial positions and the speed ratio, and they are stated in (\ref{R1R2R3}).

By (\ref{R1R2R3}), for any nonzero vector $\bm{x}\in\mathbb{R}^{n-1}$, we have
\begin{equation*}\begin{aligned}\label{positiveR1234}
\bm{x}^\mathsf{T}R_1\bm{x}=(\bm{x}^\mathsf{T}A_{12})^2+m_{12}^2\|\bm{x}\|_2^2>0.
\end{aligned}\end{equation*}
Therefore, $R_1$ is positive definite, implying that there exists a nonsingular matrix $L\in\mathbb{R}^{(n-1)\times (n-1)}$ such that $R_1=LL^\mathsf{T}$. Define $\tilde{\bm{z}}_{-n}=L^\mathsf{T}\bm{z}_{-n}$, and thus $\bm{z}_{-n}=L^{-\mathsf{T}}\tilde{\bm{z}}_{-n}$. Hence, it follows from (\ref{zminusneuqtionless1}) that $\tilde{\bm{z}}_{-n}$ satisfies the constraint \begin{equation}\begin{aligned}\label{tildezconstraint}
&\tilde{\bm{z}}_{-n}^\mathsf{T}L^{-1}R_1L^{-\mathsf{T}}\tilde{\bm{z}}_{-n}+2R_2^\mathsf{T}L^{-\mathsf{T}}\tilde{\bm{z}}_{-n}+r_3=0\\
&\Rightarrow \|\tilde{\bm{z}}_{-n}\|_2^2+2R_2^\mathsf{T}L^{-\mathsf{T}}\tilde{\bm{z}}_{-n}+r_3=0\\
&\Rightarrow\|\tilde{\bm{z}}_{-n}+L^{-1}R_2\|_2^2=R_2^\mathsf{T}R_1^{-1}R_2-r_3.
\end{aligned}\end{equation}

Note that $R_2^\mathsf{T}R_1^{-1}R_2-r_3>0$ must hold, as (\ref{tildezconstraint}) comes from the BAS (\ref{BAS1v1less1new}) with $i=1$. Thus, according to (\ref{znfromzminusn}) and (\ref{tildezconstraint}), the optimization problem (\ref{oriminimizationless1}) can be equivalently reformulated as the following problem which has only one equality constraint
\begin{equation}\begin{aligned}\label{newminimizationproblemless1}
&\min_{\tilde{\bm{z}}_{-n}\in\mathbb{R}^{n-1}}\frac{w_{12}-A_{12}^\mathsf{T}L^{-\mathsf{T}}\tilde{\bm{z}}_{-n}}{m_{12}}\\
&\, \text{s.t. }\|\tilde{\bm{z}}_{-n}+L^{-1}R_2\|_2^2=R_2^\mathsf{T}R_1^{-1}R_2-r_3.
\end{aligned}\end{equation}
Moreover, according to the transformation $\bm{z}_{-n}=L^{-\mathsf{T}}\tilde{\bm{z}}_{-n}$ and (\ref{znfromzminusn}), if $\tilde{\bm{p}}^*_{-n}$ is the solution of the problem (\ref{newminimizationproblemless1}), then the OTP $\bm{p}^*$ which is the solution of (\ref{oriminimizationless1}), is given by
\begin{equation}\label{newcoordiOTPless1}
\bm{p}^*_{-n}=L^{-\mathsf{T}}\tilde{\bm{p}}^*_{-n},p_n^*=\frac{w_{12}-A_{12}^\mathsf{T}\bm{p}_{-n}^*}{m_{12}}.
\end{equation}

Next, we focus on the solution of the problem (\ref{newminimizationproblemless1}). The associated Hamiltonian function is
\begin{equation*}\begin{aligned}\label{Hamiltonnew2v1less1}
H_2(\tilde{\bm{z}}_{-n},&\lambda)=\frac{w_{12}-A_{12}^\mathsf{T}L^{-\mathsf{T}}\tilde{\bm{z}}_{-n}}{m_{12}}\\
&+\lambda(\|\tilde{\bm{z}}_{-n}+L^{-1}R_2\|_2^2-R_2^\mathsf{T}R_1^{-1}R_2+r_3)
\end{aligned}\end{equation*}
where $\lambda\in\mathbb{R}$ is the Lagrangian multiplier. Thus, $\tilde{\bm{p}}^*_{-n}$ satisfies the following optimality conditions
\begin{subequations}\label{Optimalitynew2v1less1}\begin{align}
\frac{\partial H_2}{\partial\tilde{\bm{z}}_{-n}}&=-\frac{L^{-1}A_{12}}{m_{12}}+2(\tilde{\bm{p}}^*_{-n}+L^{-1}R_2)\lambda=0\label{Optimalitynew2v1less11}\\
\frac{\partial H_2}{\partial\lambda}&=\|\tilde{\bm{p}}^*_{-n}+L^{-1}R_2\|_2^2-R_2^\mathsf{T}R_1^{-1}R_2+r_3=0.\label{Optimalitynew2v1less12}
\end{align}\end{subequations}
Then, multiplying (\ref{Optimalitynew2v1less11}) by $(\tilde{\bm{p}}^*_{-n}+L^{-1}R_2)^\mathsf{T}$ from the left and employing the equality (\ref{Optimalitynew2v1less12}), we can obtain
\begin{equation}\label{lambda2}
\lambda=\frac{(\tilde{\bm{p}}^*_{-n}+L^{-1}R_2)^\mathsf{T}L^{-1}A_{12}}{2m_{12}(R_2^\mathsf{T}R_1^{-1}R_2-r_3)}.
\end{equation}

Note that $m_{12}$ and $\lambda$ are both scalar, and $L^{-1}A_{12}$ is not a zero vector in this case. Thus it follows from (\ref{Optimalitynew2v1less11}) that 
\begin{equation}\label{parallelm12lambda}
L^{-1}A_{12}||(\tilde{\bm{p}}^*_{-n}+L^{-1}R_2)\Rightarrow\tilde{\bm{p}}^*_{-n}+L^{-1}R_2=kL^{-1}A_{12}
\end{equation}
holds for a constant $k\in\mathbb{R}$.

Substituting (\ref{lambda2}) into (\ref{Optimalitynew2v1less11}) and combining (\ref{parallelm12lambda}) yield
\begin{equation}\begin{aligned}\label{solvefortildepn1}
(k^2\|L^{-1}A_{12}\|_2^2-R_2^\mathsf{T}R_1^{-1}R_2+r_3)L^{-1}A_{12}=0
\end{aligned}\end{equation}
and since $L^{-1}A_{12}$ is a nonzero vector, (\ref{solvefortildepn1}) leads to
\begin{equation}\label{finaltildek}\begin{aligned}
k=\sqrt{\frac{R_2^\mathsf{T}R_1^{-1}R_2-r_3}{\|L^{-1}A_{12}\|_2^2}}=\sqrt{\frac{R_2^\mathsf{T}R_1^{-1}R_2-r_3}{A_{12}^\mathsf{T}R_1^{-1}A_{12}}}.
\end{aligned}\end{equation}
Thus, according to (\ref{newcoordiOTPless1}), (\ref{parallelm12lambda}) and (\ref{finaltildek}), it can be verified that the OTP $\bm{p}^*$ is given by
\begin{equation}\begin{aligned}\label{DDSless1twoDOTP12233}
\bm{p}^*_{-n}&=\sqrt{\frac{R_2^\mathsf{T}R_1^{-1}R_2-r_3}{A_{12}^\mathsf{T}R_1^{-1}A_{12}}}R_1^{-1}A_{12}-R_1^{-1}R_2\\
p^*_n&=\frac{w_{12}-A_{12}^\mathsf{T}\bm{p}^*_{-n}}{m_{12}}.
\end{aligned}\end{equation}

Since $m_{12}>0$ and $R_1^{-1}$ is positive definite, the reason why the sign of $k$ in (\ref{finaltildek}) is positive, can be obtained by noting that $p_n^*$ in (\ref{DDSless1twoDOTP12233}) should take the minimal value. 

Case 2: Next, we consider the case $m_{12}=0$. Then, the equality (\ref{intersectionbar}), which is the difference between two constraints in (\ref{oriminimizationless1}) (equivalent to (\ref{intersectionbar123})), becomes 
\begin{equation*}\label{OTPDDSless1m12=02233}
A_{12}^\mathsf{T}\bm{z}_{-n}-w_{12}=0.
\end{equation*}
Thus, two constraints for the problem (\ref{oriminimizationless1}) can be replaced by\begin{equation}\label{newcOTPDDSless1m12=02233}
\|\bm{z}-\theta_1\|_2=\delta_1,A_{12}^\mathsf{T}\bm{z}_{-n}-w_{12}=0.
\end{equation}
The first constraint in (\ref{newcOTPDDSless1m12=02233}) can also be rewritten as
\begin{equation}\begin{aligned}\label{BAS1v1less1newd1}
(z_n-\theta_{1,n})^2=\delta_1^2-\|\bm{z}_{-n}-\theta_{1,-n}\|_2^2.\\
\end{aligned}\end{equation}

Notice that the goal of the problem (\ref{oriminimizationless1}) is to seek the minimum of $z_n$. Thus, it follows from (\ref{newcOTPDDSless1m12=02233}) and (\ref{BAS1v1less1newd1}) that the OTP $\bm{p}^*$ for the problem (\ref{oriminimizationless1}) satisfies
\begin{equation}\label{OTPDDSless1m12=0}
p_{n}^*=\theta_{1,n}-\sqrt{\delta_1^2-\|\bm{p}_{-n}^*-\theta_{1,-n}\|_2^2}
\end{equation}
and $\bm{p}_{-n}^*$ is the solution of the maximization problem\begin{equation}\label{maximizationless1m12=0}
\max_{\bm{z}_{-n}\in\mathbb{R}^{n-1}}\delta_1^2-\|\bm{z}_{-n}-\theta_{1,-n}\|_2^2,\text{ s.t. }A_{12}^\mathsf{T}\bm{z}_{-n}-w_{12}=0.
\end{equation}

The Hamiltonian function for the problem (\ref{maximizationless1m12=0}) is
\begin{equation*}\label{Hamiltonfunction3}
H_3(\bm{z}_{-n},\lambda)=\delta_1^2-\|\bm{z}_{-n}-\theta_{1,-n}\|_2^2+\lambda(A_{12}^\mathsf{T}\bm{z}_{-n}-w_{12})
\end{equation*}
and $\bm{p}^*_{-n}$ satisfies the related optimality conditions as follows:\begin{subequations}\label{Optimalitynew2v1less1hamilton3}\begin{align}
\frac{\partial H_3}{\partial\bm{z}_{-n}}&=-2(\bm{p}_{-n}^*-\theta_{1,-n})+A_{12}\lambda =0\label{Optimalitynew2v1less1hamilton31}\\
\frac{\partial H_3}{\partial\lambda}&=A_{12}^\mathsf{T}\bm{p}^*_{-n}-w_{12}=0.\label{Optimalitynew2v1less1hamilton32}
\end{align}\end{subequations}

Then, multiplying (\ref{Optimalitynew2v1less1hamilton31}) by $A_{12}^\mathsf{T}$ from the left and employing the equality (\ref{Optimalitynew2v1less1hamilton32}), we can obtain 
\begin{equation}\label{lamda22bb}
\lambda=\frac{2A_{12}^\mathsf{T}(\bm{p}_{-n}^*-\theta_{1,-n})}{\|A_{12}\|_2^2}=\frac{2(w_{12}-A_{12}^\mathsf{T}\theta_{1,-n})}{\|A_{12}\|_2^2}.
\end{equation}
Substituting (\ref{lamda22bb}) into (\ref{Optimalitynew2v1less1hamilton31}) yields 
\begin{equation}\begin{aligned}\label{BAS1v1lessyeilsds}
\bm{p}^*_{-n}&=\theta_{1,-n}+\frac{A_{12}(w_{12}-A_{12}^\mathsf{T}\theta_{1,-n})}{\|A_{12}\|_2^2}\\
&=\frac{A_{12}w_{12}+C_{12}\theta_{1,-n}}{\|A_{12}\|_2^2}.
\end{aligned}\end{equation}
Thus, according to (\ref{OTPDDSless1m12=0}) and (\ref{BAS1v1lessyeilsds}), the OTP $\bm{p}^*$ is obtained as (\ref{DDSless1twoDOTP2}) shows. Therefore, we finish the proof.
\qed

\begin{figure}\centering
\graphicspath{{figures/}}
\includegraphics[width=70mm,height=47mm]{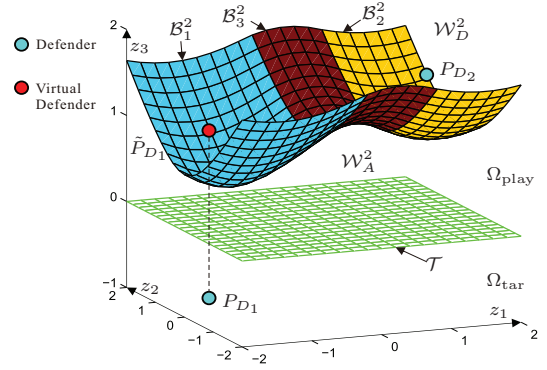}
\put(-40,125){$\scriptstyle{\mathcal{W}_D^2}$}
\put(-75,75){$\scriptstyle{\mathcal{W}_A^2}$}
\put(-20,70){$\scriptstyle{\Omega_{\rm play}}$}
\put(-20,30){$\scriptstyle{\Omega_{\rm tar}}$}
\put(-120,22){$\scriptstyle{P_{D_1}}$}
\put(-155,80){$\scriptstyle{\tilde{P}_{D_1}}$}
\put(-38,110){$\scriptstyle{P_{D_2}}$}
\put(-19,19){$\scriptstyle{z_1}$}
\put(-151,28){$\scriptstyle{z_2}$}
\put(-155,119){$\scriptstyle{z_3}$}
\put(-140,125){$\scriptstyle{\mathcal{B}^2_1}$}
\put(-120,130){$\scriptstyle{\mathcal{B}^2_3}$}
\put(-67,131){$\scriptstyle{\mathcal{B}^2_2}$}
\put(-43,36){$\scriptstyle{\mathcal{T}}$}
\caption{Barrier and winning subspaces for two defenders and one attacker with $\alpha=0.5$ in three-dimensional space ($n=3$). The barrier $\mathcal{B}^2$ consists of three parts $\mathcal{B}_1^2,\mathcal{B}_2^2$ and $\mathcal{B}_3^2$, where $\mathcal{B}_i^2(i=1,2)$ is the part only related to $P_{D_i}$, and $\mathcal{B}_3^2$ is determined by both two defenders. The barrier $\mathcal{B}^2$ divides $\Omega_{\rm play}$ into two subspaces: defender winning subspace (DWS) $\mathcal{W}_D^2$ and attacker winning subspace (AWS) $\mathcal{W}_A^2$. If $P_A$ lies in $\mathcal{W}_D^2$, two defenders can capture $P_A$ in $\Omega_{\rm play}$, and if $P_A$ lies in $\mathcal{W}_A^2$, he can reach $\Omega_{\rm tar}$ without being captured, regardless of two defenders' strategies. If $P_A$ lies at $\mathcal{B}^2$, under the optimal plays of all players, $P_A$ will be captured exactly when reaching $\mathcal{T}$.\label{fig:1}}
\end{figure}

\section{Three-Dimensional Illustrative Example}\label{threeexample}
For a better illustration of the above achievements, a subspace (half-space) guarding game in three-dimensional space is presented in this section, as Fig. \ref{fig:game} shows.

Take $n=3$, and the TH, target subspace and play subspace are given by $\mathcal{T}=\{\bm{z}\in\mathbb{R}^3|z_3=0\},\Omega_{\rm tar}=\{\bm{z}\in\mathbb{R}^3|z_3\leq0\}$ and $\Omega_{\rm play}=\{\bm{z}\in\mathbb{R}^3|z_3>0\}$ respectively, as Fig. \ref{fig:1} shows. The initial positions of two defenders (blue circles) are $\bm{x}_{D_1}^0=(-1.5,0,-1)^\mathsf{T}$ and $\bm{x}_{D_2}^0=(1.5,0,1.5)^\mathsf{T}$. Thus, $P_{D_1}$'s virtual defender $\tilde{P}_{D_1}$ (red circle), symmetric with $P_{D_1}$ about the TH $\mathcal{T}$, satisfies $\tilde{x}_{D_1}^0=(-1.5,0,1)^\mathsf{T}$. Take $\alpha=0.5$. 
\subsection{Barrier and winning subspaces}\label{threeDbarrier}
This part focuses on the qualitative result, namely, which team can guarantee to win the game.  

It follows from the classification conditions (\autoref{twotypeslema}) that two defenders are both active defenders, implying that two defenders both contribute to the barrier $\mathcal{B}^2$.

According to \autoref{thom2v1barrierless1}, we can directly obtain that the barrier $\mathcal{B}^2$ consists of three parts: $\mathcal{B}_1^2=\{\bm{z}\in\mathbb{R}^3|z_3=\sqrt{z_1^2/3+z_2^2/3+z_1+1},z_1<-7/32\},\mathcal{B}_2^2=\{\bm{z}\in\mathbb{R}^3|z_3=\sqrt{z_1^2/3+z_2^2/3-z_1+21/16},z_1>17/32\}$ and $\mathcal{B}_3^2=\{\bm{z}\in\mathbb{R}^3|z_3=\sqrt{-z_1^2+z_2^2/3+5z_1/12+719/768},-7/32\leq z_1\leq 17/32\}$, as Fig. \ref{fig:1} shows. Visually, in $\Omega_{\rm play}$, the subspace above $\mathcal{B}^2$ is the DWS $\mathcal{W}_D^2$, and below $\mathcal{B}^2$ is the AWS $\mathcal{W}_A^2$.

Thus, if $\bm{x}_A^0\in\mathcal{W}_D^2$, two defenders can guarantee to capture $P_A$ in $\Omega_{\rm play}$. If $\bm{x}_A^0\in\mathcal{W}_A^2$, $P_A$ can reach $\Omega_{\rm tar}$ without being captured, irrespective of two defenders' strategies. If $\bm{x}_A^0\in\mathcal{B}^2$, under three players' optimal plays, $P_A$ will be captured exactly when reaching $\mathcal{T}$, that is, no player wins the game.

\subsection{Optimal strategies}
This part considers the scenario of two teams trying to play the game with the best performance. Specifically, if $\bm{x}_A^0\in\mathcal{W}_D^2$, $J_\mathcal{T}$ in (\ref{DDSpf}) is considerd.

See \autoref{thomDDS2v1less1}. Take $\bm{x}_A^0=(0,0,2)^\mathsf{T}$ and thus as Fig. \ref{fig:1} shows, $\bm{x}_A^0\in\mathcal{W}_D^2$. It can be verified that $\theta_2-\delta_2\bm{e}_3=(-1/2,0,(13-2\sqrt{10})/6)\in\mathcal{R}_A^1(\bm{x}_A^0,\bm{x}_{D_1}^0,0.5)$. Hence, only $P_{D_2}$ is an effective defender. The optimal strategies for $P_{D_2}$  and $P_A$ are given by $\bm{d}_2^*=\phi(\bm{p}^*,\bm{x}_{D_2}^0)$ and $\bm{a}^*=\phi(\bm{p}^*,\bm{x}_A^0)$ respectively, where the OTP $\bm{p}^*$ is given by $(-1/2,0,(13-2\sqrt{10})/6)^\mathsf{T}$, while $P_{D_1}$ can take any strategy in $\mathcal{U}$. Thus, although $P_{D_1}$ and $P_{D_2}$ both work in the barrier construction, only $P_{D_2}$ is effective and contributes to the capture of $P_A$.

\section{Conclusion}\label{conclusion}
The differential game where two defenders guard a subspace from one attacker has been solved in analytical form, including the barrier construction and optimal strategy investigation. The barrier for one defender and two defender cases were both constructed, as it was demonstrated that under certain initial configurations, the barrier depends on only one of two defenders. Thus, the conditions about initial configurations to determine whether the barrier depends on both two defenders or only one of them, were given. The optimal strategies for three players in the DWS have also been investigated. If the attacker initially lies in the DWS, the capture can be guaranteed, and the optimal strategy for each player is equivalent to finding the capture point which was given explicitly. This subspace guarding game and its analysis lay the cornerstone for future and more challenging reach-avoid games with multiple defenders and attackers. There are few reach-avoid differential games whose state space dimension is more than nine, which have been solved in closed form.

\appendix[Proof of Lemma 1]
We prove $\mathcal{W}_A^2(\bm{x}_{D_i}^0,\bm{x}_{D_j}^0,\alpha)=\mathcal{W}_A^2(\tilde{\bm{x}}_{D_i}^0,\bm{x}_{D_j}^0,\alpha)$, which suffices to prove this lemma by noting that the barrier $\mathcal{B}^2$ is the separating surface between $\mathcal{W}_D^2$ and $\mathcal{W}_A^2$.

Suppose $\bm{z}\in\mathcal{W}_A^2(\bm{x}_{D_i}^0,\bm{x}_{D_j}^0,\alpha)$, and then there must exist a point $\bm{p}$ in $\mathcal{T}$ such that $\|\bm{z}-\bm{p}\|_2<\alpha\|\bm{x}_{D_k}^0-\bm{p}\|_2$ holds for $k=i,j$. Note that $\|\tilde{\bm{x}}_{D_i}^0-\bm{p}\|_2=\|\bm{x}_{D_i}^0-\bm{p}\|_2$. Thus, it can be obtained that $\|\bm{z}-\bm{p}\|_2<\alpha\|\tilde{\bm{x}}_{D_i}^0-\bm{p}\|_2$. 

From the above, we conclude that $\bm{z}\in\mathcal{W}_A^2(\tilde{\bm{x}}_{D_i}^0,\bm{x}_{D_j}^0,\alpha)$, implying that $\mathcal{W}_A^2(\bm{x}_{D_i}^0,\bm{x}_{D_j}^0,\alpha)\subset\mathcal{W}_A^2(\tilde{\bm{x}}_{D_i}^0,\bm{x}_{D_j}^0,\alpha)$. On the other side, suppose $\bm{z}\in\mathcal{W}_A^2(\tilde{\bm{x}}_{D_i}^0,\bm{x}_{D_j}^0,\alpha)$ and in the similar way, $\mathcal{W}_A^2(\tilde{\bm{x}}_{D_i}^0,\bm{x}_{D_j}^0,\alpha)\subset\mathcal{W}_A^2(\bm{x}_{D_i}^0,\bm{x}_{D_j}^0,\alpha)$ can be derived. Thus, we finish the proof.\qed
\ifCLASSOPTIONcaptionsoff
  \newpage
\fi



\bibliographystyle{IEEEtran}
\bibliography{references}

\begin{thebibliography}{10}
\providecommand{\url}[1]{#1}
\csname url@samestyle\endcsname
\providecommand{\newblock}{\relax}
\providecommand{\bibinfo}[2]{#2}
\providecommand{\BIBentrySTDinterwordspacing}{\spaceskip=0pt\relax}
\providecommand{\BIBentryALTinterwordstretchfactor}{4}
\providecommand{\BIBentryALTinterwordspacing}{\spaceskip=\fontdimen2\font plus
\BIBentryALTinterwordstretchfactor\fontdimen3\font minus
  \fontdimen4\font\relax}
\providecommand{\BIBforeignlanguage}[2]{{%
\expandafter\ifx\csname l@#1\endcsname\relax
\typeout{** WARNING: IEEEtran.bst: No hyphenation pattern has been}%
\typeout{** loaded for the language `#1'. Using the pattern for}%
\typeout{** the default language instead.}%
\else
\language=\csname l@#1\endcsname
\fi
#2}}
\providecommand{\BIBdecl}{\relax}
\BIBdecl

\bibitem{Ho1098197}
Y.~Ho, A.~Bryson, and S.~Baron, ``Differential games and optimal
  pursuit-evasion strategies,'' \emph{IEEE Transactions on Automatic Control},
  vol.~10, no.~4, pp. 385--389, Oct 1965.

\bibitem{petrosjan1993differential}
L.~A. Petrosjan, \emph{Differential games of pursuit}.\hskip 1em plus 0.5em
  minus 0.4em\relax World Scientific, 1993, vol.~2.

\bibitem{engwerda2005lq}
J.~Engwerda, \emph{LQ dynamic optimization and differential games}.\hskip 1em
  plus 0.5em minus 0.4em\relax John Wiley \& Sons, 2005.

\bibitem{Ba1999Dynamic}
T.~Basar and G.~J. Olsder, \emph{Dynamic Noncooperative Game Theory}.\hskip 1em
  plus 0.5em minus 0.4em\relax SIAM, 1999.

\bibitem{Liu6907859}
S.~Liu, Z.~Zhou, C.~Tomlin, and J.~K. Hedrick, ``Evasion of a team of dubins
  vehicles from a hidden pursuer,'' in \emph{2014 IEEE International Conference
  on Robotics and Automation (ICRA)}, May 2014, pp. 6771--6776.

\bibitem{Dong7862774}
X.~Dong and G.~Hu, ``Time-varying formation tracking for linear multiagent
  systems with multiple leaders,'' \emph{IEEE Transactions on Automatic
  Control}, vol.~62, no.~7, pp. 3658--3664, July 2017.

\bibitem{Mitchell2005A}
I.~M. Mitchell, A.~M. Bayen, and C.~J. Tomlin, ``A time-dependent
  {H}amilton-{J}acobi formulation of reachable sets for continuous dynamic
  games,'' \emph{IEEE Transactions on Automatic Control}, vol.~50, no.~7, pp.
  947--957, Jul 2005.

\bibitem{Margellos2011Ham}
K.~Margellos and J.~Lygeros, ``Hamilton-{J}acobi formulation for reach-avoid
  differential games,'' \emph{IEEE Transactions on Automatic Control}, vol.~56,
  no.~8, pp. 1849--1861, Aug 2011.

\bibitem{Fisac2015Reach}
J.~F. Fisac, M.~Chen, C.~J. Tomlin, and S.~S. Sastry, ``Reach-avoid problems
  with time-varying dynamics, targets and constraints,'' in \emph{International
  Conference on Hybrid Systems: Computation and Control}, 2015, pp. 11--20.

\bibitem{Barron2017Reach}
E.~N. Barron, ``Reach-avoid differential games with targets and obstacles
  depending on controls,'' \emph{Dynamic Games \& Applications}, no.~4, pp.
  1--17, 2017.

\bibitem{Mohanan2018Toward}
J.~Mohanan, S.~R. Manikandasriram, R.~H. Venkatesan, and B.~Bhikkaji, ``Toward
  real-time autonomous target area protection: Theory and implementation,''
  \emph{IEEE Transactions on Control Systems Technology}, pp. 1--8, 2018.

\bibitem{ChenMo2016Multiplayer}
M.~Chen, Z.~Zhou, and C.~J. Tomlin, ``Multiplayer reach-avoid games via
  pairwise outcomes,'' \emph{IEEE Transactions on Automatic Control}, vol.~62,
  no.~3, pp. 1451--1457, Mar 2017.

\bibitem{Garcia8340791}
E.~Garcia, D.~W. Casbeer, and M.~Pachter, ``Design and analysis of
  state-feedback optimal strategies for the differential game of active
  defense,'' \emph{IEEE Transactions on Automatic Control}, pp. 1--1, 2018.

\bibitem{Mylvaganam7909033}
T.~Mylvaganam, M.~Sassano, and A.~Astolfi, ``A differential game approach to
  multi-agent collision avoidance,'' \emph{IEEE Transactions on Automatic
  Control}, vol.~62, no.~8, pp. 4229--4235, Aug 2017.

\bibitem{Is1967Diff}
R.~Isaacs, \emph{Differential Games}.\hskip 1em plus 0.5em minus 0.4em\relax
  New York: Wiley, 1967.

\bibitem{Li5751240}
D.~Li and J.~B. Cruz, ``Defending an asset: A linear quadratic game approach,''
  \emph{IEEE Transactions on Aerospace and Electronic Systems}, vol.~47, no.~2,
  pp. 1026--1044, April 2011.

\bibitem{YanReaTwo2018}
R.~Yan, Z.~Shi, and Y.~Zhong, ``Reach-avoid games with two defenders and one
  attacker: An analytical approach,'' \emph{IEEE Transactions on Cybernetics},
  vol.~PP, no.~99, pp. 1--12, 2018.

\bibitem{Chen8267187}
M.~Chen, S.~L. Herbert, M.~S. Vashishtha, S.~Bansal, and C.~J. Tomlin,
  ``Decomposition of reachable sets and tubes for a class of nonlinear
  systems,'' \emph{IEEE Transactions on Automatic Control}, vol.~63, no.~11,
  pp. 3675--3688, Nov 2018.

\bibitem{Maidens2018Exploiting}
J.~Maidens and M.~Arcak, ``Exploiting symmetry for discrete-time reachability
  computations,'' \emph{IEEE Control Systems Letters}, vol.~2, no.~2, pp.
  213--217, April 2018.

\bibitem{Kariotoglou2017The}
N.~Kariotoglou, M.~Kamgarpour, T.~Summers, and J.~Lygeros, ``The linear
  programming approach to reach-avoid problems for markov decision processes,''
  \emph{Mathematics}, vol.~60, pp. 263--285, 2017.

\bibitem{Mo7989015}
M.~Chen, S.~Herbert, and C.~J. Tomlin, ``Exact and efficient hamilton-jacobi
  guaranteed safety analysis via system decomposition,'' in \emph{2017 IEEE
  International Conference on Robotics and Automation (ICRA)}, May 2017, pp.
  87--92.

\bibitem{Xue2017Rea}
B.~Xue, A.~Easwaran, N.~J. Cho, and M.~Fränzle, ``Reach-avoid verification for
  nonlinear systems based on boundary analysis,'' \emph{IEEE Transactions on
  Automatic Control}, vol.~62, no.~7, pp. 3518--3523, July 2017.

\bibitem{ZHOU201664}
Z.~Zhou, W.~Zhang, J.~Ding, H.~Huang, D.~M. Stipanović, and C.~J. Tomlin,
  ``Cooperative pursuit with voronoi partitions,'' \emph{Automatica}, vol.~72,
  pp. 64 -- 72, 2016.

\bibitem{Pierson2017Int}
A.~Pierson, Z.~Wang, and M.~Schwager, ``Intercepting rogue robots: An algorithm
  for capturing multiple evaders with multiple pursuers,'' \emph{IEEE Robotics
  and Automation Letters}, vol.~2, no.~2, pp. 530--537, April 2017.

\bibitem{Bakolas2012Relay}
E.~Bakolas and P.~Tsiotras, ``Relay pursuit of a maneuvering target using
  dynamic {V}oronoi diagrams,'' \emph{Automatica}, vol.~48, no.~9, pp.
  2213--2220, Sep 2012.

\bibitem{Ramana2016Pursuit}
M.~V. Ramana and M.~Kothari, ``Pursuit strategy to capture high-speed evaders
  using multiple pursuers,'' \emph{Journal of Guidance Control \& Dynamics},
  pp. 1--11, 2016.

\bibitem{Oyler2016Pursuit}
D.~W. Oyler, P.~T. Kabamba, and A.~R. Girard, ``Pursuit-evasion games in the
  presence of obstacles,'' \emph{Automatica}, vol.~65, pp. 1 -- 11, Mar 2016.

\bibitem{Eklund2012Switched}
J.~M. Eklund, J.~Sprinkle, and S.~S. Sastry, ``Switched and symmetric
  pursuit/evasion games using online model predictive control with application
  to autonomous aircraft,'' \emph{IEEE Transactions on Control Systems
  Technology}, vol.~20, no.~3, pp. 604--620, May 2012.

\bibitem{Polak2016Method}
E.~Polak, S.~Lee, I.~Bustany, and A.~Madhan, ``Method of outer approximations
  and adaptive approximations for a class of matrix games,'' \emph{Journal of
  Optimization Theory \& Applications}, vol. 170, no.~3, pp. 1--24, 2016.

\bibitem{Raslan7490540}
H.~Raslan, H.~Schwartz, and S.~Givigi, ``A learning invader for the guarding a
  territory game,'' in \emph{2016 Annual IEEE Systems Conference (SysCon)},
  April 2016, pp. 1--8.

\bibitem{Shinar1102372}
J.~Shinar and S.~Gutman, ``Three-dimensional optimal pursuit and evasion with
  bounded controls,'' \emph{IEEE Transactions on Automatic Control}, vol.~25,
  no.~3, pp. 492--496, Jun 1980.

\bibitem{Li7482692}
W.~Li, ``A dynamics perspective of pursuit-evasion: Capturing and escaping when
  the pursuer runs faster than the agile evader,'' \emph{IEEE Transactions on
  Automatic Control}, vol.~62, no.~1, pp. 451--457, Jan 2017.

\bibitem{HAYOUN2017122}
S.~Y. Hayoun and T.~Shima, ``On guaranteeing point capture in linear n-on-1
  endgame interception engagements with bounded controls,'' \emph{Automatica},
  vol.~85, pp. 122 -- 128, 2017.

\bibitem{Bopardikar20112067}
S.~D. Bopardikar, S.~L. Smith, and F.~Bullo, ``On vehicle placement to
  intercept moving targets,'' \emph{Automatica}, vol.~47, no.~9, pp. 2067 --
  2074, Sep 2011.

\bibitem{merz1971homicidal}
A.~W. Merz, ``The homicidal chauffeur-a differential game,'' Stanford
  University, Tech. Rep., 1971.

\bibitem{getz1981two}
W.~Getz and M.~Pachter, ``Two-target pursuit-evasion differential games in the
  plane,'' \emph{Journal of Optimization Theory and Applications}, vol.~34,
  no.~3, pp. 383--403, 1981.

\bibitem{ruiz2016differential}
U.~Ruiz and R.~Murrieta-Cid, ``A differential pursuit/evasion game of capture
  between an omnidirectional agent and a differential drive robot, and their
  winning roles,'' \emph{International Journal of Control}, vol.~89, no.~11,
  pp. 2169--2184, 2016.

\bibitem{Sun2015Pur}
W.~Sun and P.~Tsiotras, ``Pursuit evasion game of two players under an external
  flow field,'' in \emph{2015 American Control Conference (ACC)}, Jul 2015, pp.
  5617--5622.

\bibitem{Lsler5354443}
N.~Karnad and V.~Isler, ``Lion and man game in the presence of a circular
  obstacle,'' in \emph{2009 IEEE/RSJ International Conference on Intelligent
  Robots and Systems}, Oct 2009, pp. 5045--5050.

\bibitem{Zha2016Con}
W.~Zha, J.~Chen, Z.~Peng, and D.~Gu, ``Construction of barrier in a fishing
  game with point capture,'' \emph{IEEE Transactions on Cybernetics}, vol.~PP,
  no.~99, pp. 1--14, Jun 2016.

\bibitem{Bhattacharya2016}
S.~Bhattacharya, T.~Ba{\c{s}}ar, and N.~Hovakimyan, ``A visibility-based
  pursuit-evasion game with a circular obstacle,'' \emph{Journal of
  Optimization Theory and Applications}, vol. 171, no.~3, pp. 1071--1082, Aug
  2016.

\bibitem{MACIAS2018271}
V.~Macias, I.~Becerra, R.~Murrieta-Cid, H.~M. Becerra, and S.~Hutchinson,
  ``Image feedback based optimal control and the value of information in a
  differential game,'' \emph{Automatica}, vol.~90, pp. 271 -- 285, 2018.

\bibitem{Elliott1972The}
R.~J. Elliott and N.~J. Kalton, ``The existence of value in differential
  games,'' \emph{Memoirs of the American Mathematical Society}, vol. 126, no.
  126, pp. 504--523, 1972.

\end{thebibliography}

\vfill
%

%




\vfill


\end{document}